\documentclass[aps,prl,reprint,superscriptaddress,nofootinbib,10pt]{revtex4-2}

\usepackage{graphicx,amsmath,amssymb,dcolumn,physics,color}

\newcommand{\mbr}{\ensuremath{\mathbf{r}}}

\newcommand{\bsk}{\ensuremath{\boldsymbol{{k}}}}
\newcommand{\bsr}{\ensuremath{\boldsymbol{{r}}}}
\newcommand{\dbsr}{\ensuremath{\dd[2]{\bs{r}}}}

\newcommand{\bs}[1]{\ensuremath{\boldsymbol{#1}}}

\graphicspath{{./figs/}{./}}

\begin{document}

\title{
    Pattern formation of phase-separated lipid domains in bilayer membranes
}

\author{Qiwei Yu}
\affiliation{Department of Mechanical and Aerospace Engineering, Princeton University, Princeton, NJ 08544}
\affiliation{Lewis-Sigler Institute for Integrative Genomics, Princeton University, Princeton, NJ 08544}

\author{Andrej Ko\v{s}mrlj}
\affiliation{Department of Mechanical and Aerospace Engineering, Princeton University, Princeton, NJ 08544}
\affiliation{Princeton Materials Institute, Princeton University, Princeton, NJ 08544}

\date{\today}

\begin{abstract}
    Giant unilamellar vesicles (GUVs) composed of as few as three lipid species can phase separate into small-scale lipid domains with stripes and dots patterns. These patterns have been experimentally characterized in terms of how their size and morphology depend on temperature, membrane composition, and surface tension, which revealed inconsistencies with existing theoretical models. Here, we demonstrate that the experiments can be explained with a theory that considers both the elastic deformation of the membrane and the phase separation of lipids, which are coupled by a preferred bilayer curvature. We combine analytical and numerical approaches to elucidate how characteristic pattern size and morphology emerge from these interactions. The results agree with existing experiments and offer testable predictions such as non-monotonic dependence of the domain size on osmotic pressure and pattern hysteresis upon cycling external stimuli. 
    These predictions motivate new directions for understanding the spatial patterning and organization mechanisms of biological membranes.
\end{abstract}

\maketitle

\textbf{Introduction}~~
The heterogeneous spatial organization of lipids, proteins, and other components of biological membranes plays a crucial role in essential physiological functions~\cite{semrau_membrane_2009,sych_how_2021}.  
To understand the mechanism underpinning the formation of these spatial structures, several model systems have been established, including cell-derived membranes~\cite{baumgart_large-scale_2007} and artificial membrane systems such as giant unilamellar vesicles (GUVs)~\cite{dietrich_lipid_2001,dimova_practical_2006}. 
Model membranes consisting of multiple lipid species can undergo phase separation, which organizes the membrane into domains with distinct compositions. These phase-separated domains participate in the regulation of various physiological processes, such as signaling, immune response, and vesicle trafficking~\cite{sengupta_lipid_2007,kabouridis_lipid_2006,manes_lipid_2006,parton_lipid_2003,salaun_lipid_2004}. 
Notably, phase separation in yeast vacuoles can promote survival under stress by organizing proteins for a nutrient-sensing pathway or by facilitating lipophagy~\cite{murley_sterol_2017,seo_ampk_2017,leveille_yeast_2022}.

Macroscopic liquid-liquid phase separation has been observed in artificial membranes consisting of as few as three components~\cite{dietrich_lipid_2001,veatch_organization_2002,veatch_separation_2003}.
When the membrane is quenched into the 2-phase region, phase separation commences with the nucleation of many small droplets, which subsequently grow in size until reaching the dimension of the system~\cite{stanich_coarsening_2013}.
Eventually, only a few macroscopic domains persist. 
Under certain conditions (such as establishing bilayer asymmetry by introducing excess area to the outer leaflet~\cite{cornell_tuning_2018}), however, the domains can remain at a small size and form stable periodic patterns such as stripes or dots. 
In a recent experiment~\cite{cornell_tuning_2018}, these domains have been studied in detail by measuring their length scale and spatial organization under different experimental conditions, including temperature, membrane composition, and tension. 
The domain size was found to increase with surface tension but decrease with temperature, and the morphology could switch between stripes and dots as the average lipid composition was varied. 
These quantitative measurements enabled close comparisons with multiple existing theories~\cite{schmid_physical_2017}, including those considering spontaneous curvature at the level of monolayer~\cite{schick_strongly_2018,allender_theoretical_2022} and bilayer~\cite{harden_budding_2005}. Both theories captured many aspects of the experiments, but neither appeared consistent with all the observations~\cite{cornell_tuning_2018}: the monolayer theory incorrectly predicts anti-registration between leaflets; the bilayer theory does not capture the morphology near the critical points, possibly because it employed a simplified depiction of phase separation, which only contained a line tension term in the free energy. 
Other mechanisms were also considered but ruled out, including lipid dipole repulsion~\cite{usery_line_2017}, correlated critical fluctuations~\cite{honerkamp-smith_experimental_2012}, and lineactants~\cite{hirose_concentration_2012}. So far, these experiments remain unexplained. 

\begin{figure*}[tb]
    \includegraphics[width=\linewidth]{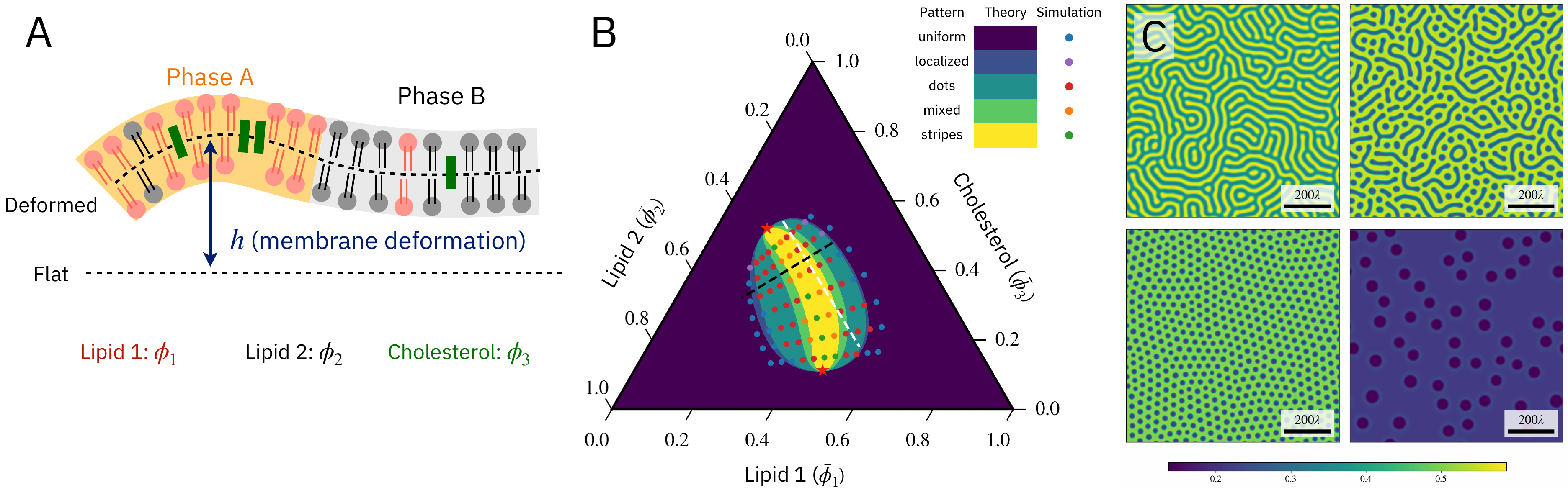}
    \caption{
        The theoretical model captures pattern formation on the membrane.
        (A) Schematics of the model. The membrane consists of three molecules: the two phospholipids (pink and gray) and cholesterol (green). They demix into phase A (yellow) and phase B (gray). Phase A is rich in lipid species 1 and therefore more curved. $h$ describes the deformation of the membrane.
        (B) Pattern morphology as a function of the average volume fractions  $\bar{\phi}_{1,2,3}$. The colored regions indicate analytical predictions of the morphology, and the colored dots are numerical simulations. The black dashed line indicates a typical tie line, with the white line indicating its perpendicular direction. The two critical points are marked by red $\star$.
        (C) Examples of the steady-state solution $\phi_1(x,y)$ with different pattern morphologies, including stripes [upper left, $(\bar{\phi_1},\bar{\phi_2})=(0.45,0.40)$], mixed [upper right, $(\bar{\phi_1},\bar{\phi_2})=(0.45,0.34)$], dots [lower left, $(\bar{\phi_1},\bar{\phi_2})=(0.43,0.29)$], and localized [lower right, $(\bar{\phi_1},\bar{\phi_2})=(0.21,0.25)$] patterns.
        Parameters: $\chi_{12} = 1.8$, $\chi_{13} = 1.4$, $\chi_{23}=1.65$, $\xi=4.8$, $\kappa= 20$, $\sigma=0.1$, $c_1=1/\sqrt{20}=0.22$.
    }
    \label{Fig1:model}
\end{figure*}

In this work, we present a minimal theoretical model capable of capturing the existing experiments. 
The model incorporates both the elastic deformation of the membrane and the phase separation of the lipids, which are coupled by a preferred bilayer curvature. 
The preferred curvature depends on the composition of lipids and emerges from a leaflet asymmetry resulting from introducing extra lipids to the outer leaflet, which was an essential procedure for stabilizing small domains~\cite{cornell_tuning_2018}.
The coupling between curvature and composition arrests the coarsening of phase-separated domains, leading to patterns at a finite length scale. The morphology and size of the patterns are determined using amplitude expansion and confirmed by numerical simulations. 
Their dependence on the membrane composition and experimental conditions such as temperature and tension are consistent with the experimental findings.
Importantly, our model generates several predictions that can be tested in future experiments.
For example, the model predicts a non-monotonic relation between the domain size and the membrane tension, which can be examined by expanding the measurements to smaller membrane tensions. 
A phase diagram of pattern morphologies revealed rich structures which can be explored experimentally by varying the membrane composition.
Our model also predicts the hysteresis of patterns with respect to experimentally accessible parameters such as osmotic pressure and temperature. 
These results uncover new directions for probing the mechanism underlying small domains in biological membranes and provide valuable insights for understanding pattern formation on curved surfaces.

\textbf{Model.}~~The theory considers three interacting lipid species that can diffuse on a deformable bilayer membrane (Fig.~\ref{Fig1:model}A). They include two phospholipids (red and gray) (such as DPPC and DiPhyPC used in ref.~\cite{cornell_tuning_2018}) and cholesterol (green). 
The shape of the membrane is described by the height field $h(\mbr)$, defined as the distance from a flat reference plane.
The composition of the membrane is defined by the (local) volume fractions occupied by different molecules $\phi_i(\mbr)$, with $i=1,2,3$ labeling the three lipid species, respectively. They are constrained by the incompressibility condition $\sum_{i=1}^3 \phi_i(\mbr) = 1$. 
Motivated by experimental evidence~\cite{cornell_tuning_2018}, the two leaflets of the bilayer are assumed to be in registration, i.e. with the same volume fractions. Therefore, only a single set of fields $\phi_i(\mbr)$ is considered for both leaflets.
The membrane composition and shape undergo dynamics governed by the following free energy:
\begin{align}
    \mathcal{F}[h, \phi_1, \phi_2] = \int \dd[2]{\bsr} \left[ f_{e}(h, \phi_1, \phi_2) + f_{c}(\phi_1, \phi_2) \right],
\end{align}
where $f_{e}$ and $f_{c}$ are the free energy densities for the elastic deformation of the membrane and the chemical interactions between lipids, respectively.
Here, we take the small curvature approximation ($\abs{\nabla h} \ll 1$) so that all the fields can be defined on the flat reference plane [$\bsr = (x, y)$].
The free energy due to membrane deformation is given by
\begin{align}
    f_{e}(h, \phi_1, \phi_2) = \frac{\kappa}{2} \qty(\nabla^2 h- c_1\phi_1)^2 + \frac{\sigma}{2} (\nabla h)^2,
\end{align}
where $\kappa$ and $\sigma$ are the bending rigidity and surface tension of the membrane, respectively. $c_1$ is the preferred bilayer curvature induced by lipid species 1. 
This term is related to the bilayer asymmetry introduced in the experiments~\cite{cornell_tuning_2018}.
For a single leaflet, preferred curvature can arise from the non-cylindrical shapes of the lipid molecules. For a bilayer, however, opposite curvatures are induced in the two leaflets, which completely cancel out unless there is an asymmetry between the two leaflets. 
Indeed, the observed patterns of small domains only emerged after extra lipids (DPPC) were introduced to the outer leaflet~\cite{cornell_tuning_2018}, which creates an asymmetry between the two leaflets and enables a non-zero preferred bilayer curvature. This suggests that the preferred curvature plays an important role in pattern formation. 
The value of $c_1$ is related to the amount of asymmetry introduced to the bilayer, which can be controlled by the type and amount of lipids added to the outer leaflet.
Although all lipid species can in principle induce preferred curvature, we only consider lipid species 1 for simplicity and for consistency with experiments~\cite{cornell_tuning_2018} where bilayer asymmetry is only introduced in the DPPC composition. Generalization to multiple curvature-generating lipids is straightforward by replacing $c_1\phi_1$ with $\sum_i c_i\phi_i$.

The second part of the free energy comes from the interaction and mixing of lipids:
\begin{align}
    &f_{c}(\phi_1, \phi_2)          = nk_BT\left[\sum_{i=1}^3 \phi_i \ln \phi_i    \right.  \nonumber                                         \\
     &\quad  \left.+ \frac{1}{2}\sum_{i,j=1}^3 \chi_{i,j}  \qty(\phi_i\phi_j - \lambda^2 \nabla\phi_i\cdot\nabla\phi_j) + \xi \phi_1\phi_2\phi_3\right],
\end{align}
where $\phi_3=1-\phi_1-\phi_2$. The free energy is an extension of the Flory-Huggins model for regular solutions~\cite{flory_thermodynamics_1942,huggins_solutions_1941}, with $n$ being the number density of the lipids and $k_B$ the Boltzmann constant. $\chi_{i,j}$ is the two-body interaction parameter between $\phi_i$ and $\phi_j$ with interaction range $\lambda$. The term containing gradients $\nabla \phi_i$ describes the interfacial properties with characteristic interface width given by $\lambda$. 
$\xi$ characterizes the three-body interaction, which was shown to be important for the closed-loop miscibility gap~\cite{idema_phase_2009} reported for these membranes~\cite{veatch_closed-loop_2006,cornell_tuning_2018}, where the presence of all three components is required for phase separation to occur.

The free energy governs the time evolution of the fields, with $h$ undergoing unconserved (model A) dynamics and $\phi_{1,2}$ undergoing conserved (model B) dynamics~\cite{hohenberg_theory_1977}:
\begin{align}\label{Eq:dynamics}
    \pdv{h}{t} = - M_h \pdv{\mathcal{F}}{h}, \quad \pdv{\phi_i}{t} = M_i \nabla^2 \pdv{\mathcal{F}}{\phi_i}\ (i=1,2),
\end{align}
where $M_h$ and $M_i$ are the mobilities of $h$ and $\phi_i$, respectively.
$\phi_3=1-\phi_1-\phi_2$ is fully determined by $\phi_1$ and $\phi_2$, and its dynamics is not considered explicitly. 
We measure energy in units of $nk_BT$ and consider the nondimensionalzed free energy density due to chemical interactions $\bar{f}_c=f_c/(n k_BT)$ as well as the nondimensionalized material properties $\bar{\kappa}=\kappa/(nk_BT)$, $\bar{\sigma}=\sigma\lambda^2/(nk_BT)$. Space is rescaled by $\bar{x}=x/\lambda$, and the preferred curvature is rescaled by $\bar{c}_1 = \lambda c_1$. For the sake of simplicity, we assume equal mobilities $M_1=M_2=M$ and rescale time by $\bar{t}=t/t_0=t n k_BTM/\lambda^2$ so that the rescaled mobilities are $1$. 
$h$ is nondimensionalized so that the rescaled $M_h$ is also $1$. 
From now on, We will omit $\bar{\ }$ and refer to nondimensionlaize quantities unless explicitly stated.

In this theory, the preferred curvature plays the key role of coupling deformation $h$ and concentrations $\phi_i$. In the absence of preferred curvature ($c_1=0$), the membrane is flat ($h=0$) and the concentrations behave like a typical ternary liquid mixture, which is either uniform or phase-separated. When phase-separated, the domains coarsen perpetually until they reach the system size. These large domains have been observed in many membrane systems~\cite{veatch_separation_2003,yanagisawa_growth_2007,stanich_coarsening_2013}. 
By coupling the concentrations to membrane deformation, the preferred curvature introduces an effective free energy cost against coarsening, which, when significant enough, leads to the formation of small domains. The typical morphologies of these domains are shown in Fig.~\ref{Fig1:model}C, including stripes, dots (arranged in a hexagonal lattice), mixed states (a combination of stripes and dots), and localized states (isolated and stationary droplets). 
These patterns are similar to those observed in experiments~\cite{cornell_tuning_2018} and are reminiscent of those found in phase field crystal models~\cite{thiele_localized_2013,knobloch_localized_2016,emmerich_phase-field-crystal_2012,elder_amplitude_2010,stefanovic_phase-field_2006,matthews_pattern_2000,cox_envelope_2004}.
The stable morphology of the pattern depends on the average composition of the membrane $\bar{\phi}_i$, which is illustrated by a phase diagram (Fig.~\ref{Fig1:model}B). The parameters were chosen such that the phase diagram qualitatively resembles that measured experimentally~\cite{cornell_tuning_2018}. The same set of parameters is used throughout unless varied explicitly. 
In the following, we characterize these patterns combining analytical and numerical methods and discuss implications for both existing and future experiments.

\begin{figure}[tb]
    \includegraphics[width=\linewidth]{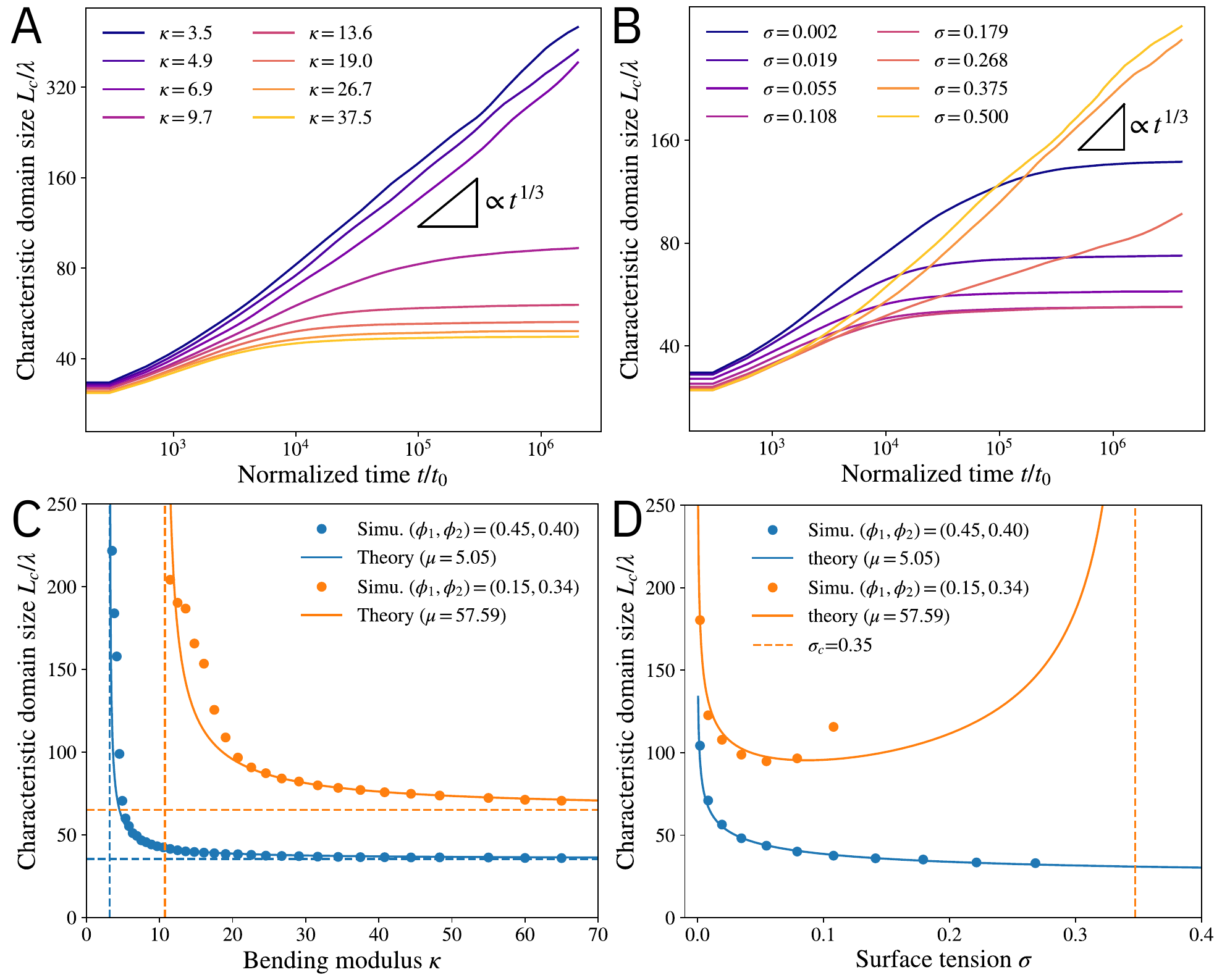}
    \caption{
        (A--B)~The time evolution of the characteristic domain size $L_c$ for different (A)~bending modulus $\kappa$ and (B)~membrane tension $\sigma$. The black triangles indicate $t^{1/3}$ scaling. Mean composition: $(\bar{\phi}_1,\bar{\phi}_2)=(0.38,0.36)$.
        Time is measured in units of $t_0=\lambda^2/(Mnk_BT)$. 
        (C)~The characteristic domain size $\tilde{L}_c = L_c/\lambda$ as a function of the bending modulus $\kappa$. The solid lines indicate theory [Eq.~\eqref{Eq:kc theory}] with the fitted value of the line tension $\mu$ between phases. The horizontal dashed lines are the limit of infinite bending modulus $\kappa\rightarrow \infty$, and the vertical dashed lines show the critical value $\kappa_c=\sqrt{\sigma\mu}/c_1$ below which the characteristic domain size becomes unbounded.
        (D)~The characteristic domain size  $\tilde{L}_c$ as a function of the membrane tension $\sigma$. The vertical dashed line shows $\sigma_c=\kappa^2c_1^2/\mu$ beyond which the characteristic domain size becomes unbounded.
        For the blue curve, $\sigma_c$ is outside the range of the plot. 
        Parameters are the same as Fig.~\ref{Fig1:model} unless otherwise stated. 
    }
    \label{Fig2:lengthscales}
\end{figure}

\textbf{Membrane deformation arrests domain coarsening.}~~We start by analyzing the size of the small domains. Fig.~\ref{Fig2:lengthscales}AB shows the time evolution of domain size $L_c(t)$ for different bending modulus $\kappa$ and membrane tension $\sigma$. In the limit of small $\kappa$ and large $\sigma$, the system behaves like a typical ternary mixture, with domain coarsening following the standard $L_c(t)\propto t^{1/3}$ scaling (black triangles). 
This scaling is consistent with both coalescence and Ostwald ripening~\cite{lifshitz_kinetics_1961,camley_dynamic_2011}, and has been reported in taut membranes~\cite{stanich_coarsening_2013}.
For large $\kappa$ and small $\sigma$, however, domain coarsening is arrested at a finite length scale, which sets the size of the patterns. 

The finite droplet size is selected by a competition between the energy costs of membrane deformation and phase separation, with the former preferring small-scale variations and the latter large domains. The deformation energy depends on the height field $h(\mbr)$, whose steady-state solution can be given in the Fourier space:
\begin{align}
    h(\bsk) = - \frac{\kappa c_1}{\kappa k^2+\sigma} \delta\phi_1(\bsk),
\end{align}
where $k=\abs{\bsk}$ and $\delta\phi_1(\bsk)$ is the Fourier transform of $\delta\phi_1(\bsr) = \phi_1(\bsr)-\bar{\phi_1}$. Therefore, the deformation field follows the concentration profile. It stays flat for the uniform state and gets modulated once patterns form (Fig.~\ref{Fig1:model}C). 
Substituting $h$ simplifies the steady-state free energy due to membrane deformation:
\begin{align}
    \mathcal{F}_{e}[\delta \phi_1]=\frac{1}{2} \int \frac{\sigma\kappa c_1^2}{\sigma+ \kappa k^2} \abs{\delta\phi_1(\bsk)}^2\dd[2]{\bsk}.
\end{align}
The coefficient $\frac{\sigma\kappa c_1^2}{\sigma+ \kappa k^2}$ decreases monotonically with $k$, thereby favoring small length scales. However, it has to compete with the Flory-Huggins free energy $\mathcal{F}_{c}$, which always favors large domains in order to reduce the interface length between phases. 
Since $\mathcal{F}_{e}$ increases monotonically with $\kappa$, it dominates over $\mathcal{F}_{c}$ when $\kappa$ is large enough, resulting in finite domains. Conversely, $\mathcal{F}_{c}$ dominates in the small $\kappa$ limit, leading to $t^{1/3}$ scaling as shown in Fig.~\ref{Fig2:lengthscales}A.

In the absence of membrane deformation energy ($f_e$), phase separation occurs along tie lines of the ternary mixture phase diagram (determined by $f_c$), which can be extracted from the convex hull of the free energy landscape~\cite{mao_phase_2019}. Motivated by numerical evidence, we assume that the full system ($h\neq 0$) also phase-separates along the same tie lines, whose slope gives the ratio of the concentration change $\delta\phi_{2}/\delta\phi_{1}$, where $\delta\phi_{i}=\phi_{i}-\bar{\phi}_{i}$ with $\bar{\phi}_{i}$ being the average composition of component $i$. The assumption fixes the ratios of $\delta\phi_i$ but not their magnitudes. Consequently, $\mathcal{F}_c$ can be simplified by expanding along the tie line. Combining it with $\mathcal{F}_{e}$ leads to an effective total free energy:
\begin{align}\label{Eq:effective_free_energy}
    \mathcal{F}_\mathrm{eff}[\delta \phi_1]=\frac{1}{2} \int \qty(\frac{\sigma\kappa c_1^2}{\sigma+ \kappa k^2} + \mu k^2) \abs{\delta\phi_1(\bsk)}^2\dd[2]{\bsk} \nonumber \\
    +\int \qty[-\frac{a}{2}\delta\phi_1(\bsr)^2 + \frac{c}{3} \delta\phi_1(\bsr)^3 + \frac{b}{4} \delta\phi_1(\bsr)^4]\dbsr,
\end{align}
where $a,b,c$ are Ginzburg-Landau coefficients and $\mu$ is related to the line tension between phases. They can be obtained by expanding the free energy along the tie lines (see SI, section IB).
The selection of the preferred length scale is dictated by the first term in the free energy,
Minimizing it with respect to $k$ leads to the characteristic length scale (domain size)
\begin{align}
    L_c = \frac{2\pi}{k_c} = \frac{2\pi}{\sqrt{\sqrt{\frac{\sigma c_1^2}{\mu}}-\frac{\sigma}{\kappa}}},
    \label{Eq:kc theory}
\end{align}
with $k_c$ being the characteristic wavevector.  

The predicted $L_c$ is confirmed by numerical solution of the dynamical equations. 
Fig.~\ref{Fig2:lengthscales}C presents $L_c$ as a function of the bending modulus $\kappa$. While small $\kappa$ always leads to perpetual coarsening ($L_c\to\infty$), coarsening is arrested beyond a critical value $\kappa_c = \sqrt{\sigma\mu}/c_1$ (vertical dashed lines). The domain size $L_c$ decreases monotonically with $\kappa$ and converges to $L_c=2\pi\qty(\sigma c_1^2/\mu)^{-1/4}$ in the infinite $\kappa$ limit (horizontal dashed lines). The theory (solid lines) agrees well with numerical simulations (circles), with a single fitting parameter $\mu$ to account for a modified line tension between the two phases %
since the chemical composition at the interface deviates slightly from the tie lines. 
$\mu$ can also be estimated by expanding the free energy along the tie lines (see SI, section IB), which leads to a slightly worse agreement (Fig.~S2).

The domain size $L_c$ has a different functional dependence on the membrane tension $\sigma$ (Fig.~\ref{Fig2:lengthscales}D). Finite domains only emerge when $\sigma<\sigma_c = \kappa^2c_1^2/\mu$. $L_c$ reaches minimum at an intermediate value $\sigma_m = \kappa^2c_1^2/(4\mu) = \sigma_c/4$ and diverges in limits $\sigma\to0$ and $\sigma\to\sigma_c$.
In the $\sigma\to 0$ limit, the energy cost for stretching is absent, so the preferred curvature condition can be satisfied everywhere on the membrane, allowing for indefinite coarsening.
In addition to the divergence of domain size, another possibility in the large $\sigma$ limit (taut membrane) is that the membrane remains flat inside each of the domains, and the system behaves like a ternary mixture with either a uniform concentration or coarsening droplets. The transition to the large $\sigma$ regime will be discussed in a later section.  
As shown in Fig.~\ref{Fig2:lengthscales}D, the predicted $L_c$ agrees well with numerical simulations.  %
The membrane tension can be varied by tuning the exterior osmotic pressure, which was found to monotonically increase domain size with tension for synthetic and cell-derived membranes~\cite{cornell_tuning_2018}. 
These experiments are likely performed at relatively high membrane tension since the vesicles appeared taut before introducing DPPC to the outer leaflet. Moreover, large domains were observed toward the end of the experiment, indicating that $\sigma$ eventually exceeds $\sigma_c$. 
Hence, they are consistent with the right branch  ($\sigma>\sigma_m$) of the predicted $L_c(\sigma)$ (Fig.~\ref{Fig2:lengthscales}D).
It will be interesting to measure the domain size at low membrane tension (which can be achieved by high exterior osmotic pressure) and look for potential non-monotonicity with tension.

Finally, we consider the effect of temperature $T$. While all the material properties and interaction parameters can depend on temperature, we assume that none of them varies drastically in the experimental range. 
Specifically, since energy is measured in units of $nk_BT$, the rescaled parameters $\bar{\kappa}$, $\bar{\sigma}$, $\chi$, and $\xi$ all have a $1/(k_BT)$ dependence on temperature, while the entropy term does not. 
Since the two-phase region in the phase diagram shrinks with increasing temperature~\cite{veatch_closed-loop_2006}, we expect the composition difference between the two phases and thus the line tension to decrease with temperature. Thus, the rescaled line tension $\mu$ will likely decrease more rapidly than $T^{-1}$. Under this assumption,  Eq.~\eqref{Eq:kc theory} predicts that the characteristic domain size $L_c$ decreases with temperature $T$, which
is confirmed by numerical simulations (see SI, section III) and is consistent with experiments~\cite{cornell_tuning_2018}.

\textbf{Pattern morphology predicted by amplitude expansion.}~~Next, we determine the steady-state pattern morphologies with an amplitude expansion around the uniform state~\cite{cross_pattern_1993}:
\begin{align} \label{Eq:amplitude expansion}
    \delta\phi_1 = \phi_1-\bar{\phi_1} = \sum_n A_{n} e^{i\bsk_n\cdot\bsr} + \mathrm{c.c.},
\end{align}
where $A_{n}$ are the amplitudes of the Fourier modes and $\mathrm{c.c.}$ stands for the complex conjugate terms. For homogeneous patterns, it suffices to consider spatially uniform amplitudes. 
The time evolution of the amplitudes can be determined by substituting Eq.~\eqref{Eq:amplitude expansion} back to the equation of motion $\pdv{\delta\phi_1}{t} = \nabla^2 \frac{\partial \mathcal{F}_\mathrm{eff}}{\partial \delta\phi_1}$ and keeping only the leading order terms in the amplitudes. 
This results in a dynamical system for $\{A_n(t)\}$, whose fixed points correspond to different pattern morphologies. The stability of these fixed points determines possible steady-state patterns (see SI, section I). Since the theory is an expansion in amplitudes, it is the most accurate near the critical points as the higher order terms in $A_n$ become less important.

We find three fixed points representing uniform, dot (Fig.~\ref{Fig1:model}C, lower left), and stripe (Fig.~\ref{Fig1:model}C, upper left) morphologies, respectively. Their stability is governed by the following control parameter:
\begin{align} \label{Eq:g definition}
    g = \frac{a_\mathrm{eff}b}{c^2} \equiv \frac{b}{c^2}\qty(a - 2  \sqrt{c_1^2\sigma\mu} + \frac{\mu\sigma}{\kappa}).
\end{align}
The uniform, dot, and stripe fixed points are stable when $g\in(-\infty,0)$,  $\qty(-\frac{1}{15}, \frac{16}{3})$, and $\qty(\frac{4}{3},+\infty)$, respectively.
Stripes and dots are bistable in the overlap region $g\in\qty(\frac{4}{3},\frac{16}{3})$; their coexistence is referred to as the mixed state (Fig.~\ref{Fig1:model}C, upper right).
Similarly, for $g\in\qty(-\frac{1}{15},0)$, individual dots are stable in a uniform background, which leads to localized states (Fig.~\ref{Fig1:model}C, lower right) similar to those found in the Swift-Hohenberg equation~\cite{thiele_localized_2013,knobloch_localized_2016}. 

The predicted phase morphologies are summarized in a phase diagram (Fig.~\ref{Fig1:model}B), which are confirmed by numerical simulation (represented by filled circles). No fitting parameters are involved when constructing the phase diagram as $(a,b,c,\mu)$ are evaluated by expanding the free energy along the tie lines. The agreement between theory and simulation is especially good near the two critical points (indicated by red stars), where the amplitude expansion is the most accurate. Moving away from the critical points, the expansion becomes less precise, but most of the morphologies are still correctly captured. The structure of the phase diagram is robust to the form of the interaction energy of lipids $f_c$, as long as it exhibits generic phase separation behaviors.

\begin{figure}[b]
    \includegraphics[width=\linewidth]{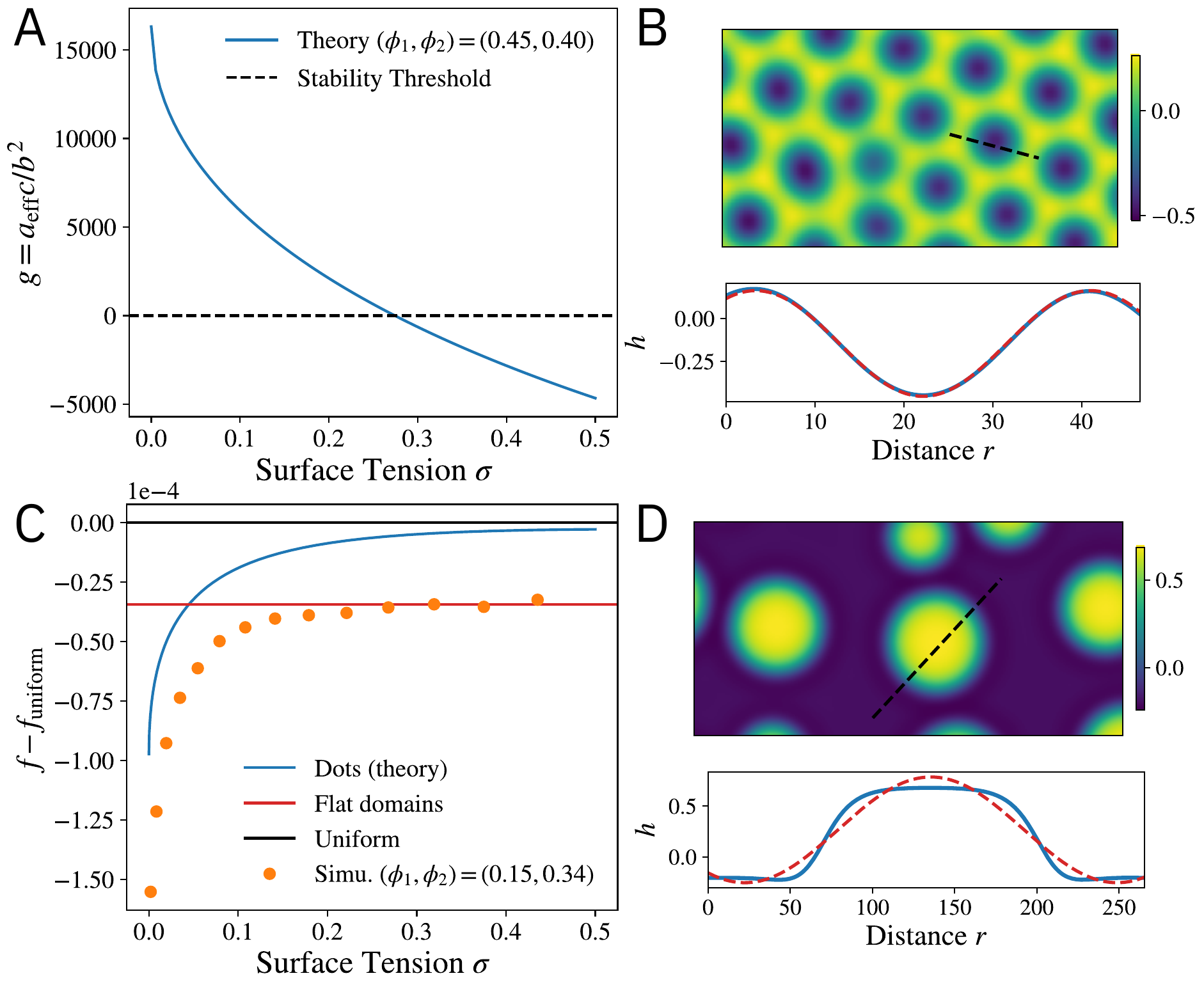}
    \caption{
        Domain coarsening in taut membranes.
        (A) The morphology control parameter $g$ as a function of $\sigma$. The black dashed line is $g=-1/15$ below which only the uniform solution is stable.
        (B) The typical deformation field $h(\bsr)$ across a droplet (blue line), which can be well described by fitting $h=h_0+h_1\sin(k r+\psi)$ (red dashed line). 
        (C)~The free energy density as a function of $\sigma$. Orange dots are simulation results; the blue line is calculated from amplitude expansion. The black and red lines are the free energy densities of the uniform state and flat droplet domains, respectively.
        (D) $h(\bsr)$ across a droplet. The deformation is flat except near the interface, which cannot be captured by fitting a shifted $\sin$ function (red dashed line).
        Composition: (A--B) $(\bar{\phi_1}, \bar{\phi_2}) = (0.45, 0.40)$; (C--D) $(\bar{\phi_1}, \bar{\phi_2}) = (0.15, 0.34)$. Parameters are the same as Fig.~\ref{Fig1:model} unless otherwise stated. 
    }
    \label{Fig3:coarsening}
\end{figure}

\begin{figure*}
    \includegraphics[width=0.9\linewidth]{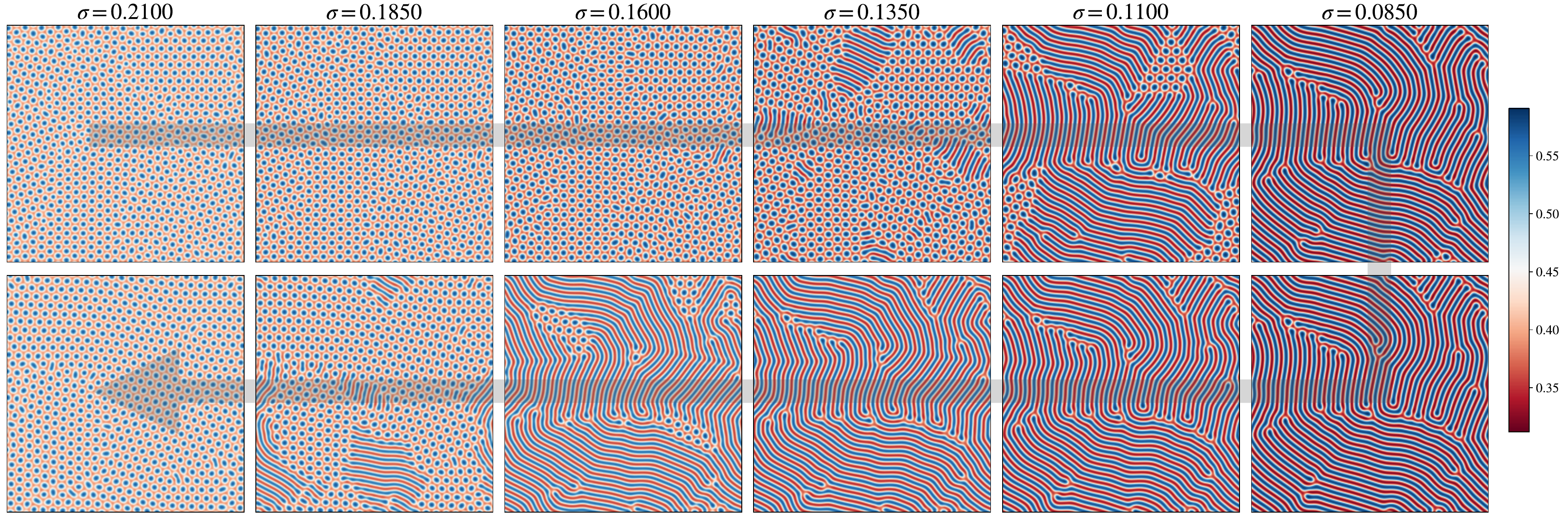}
    \caption{
        Pattern hysteresis due to tuning the surface tension $\sigma$, which is first decreased and then increased (as indicated by the arrow), with a fixed composition $(\bar{\phi_1},\bar{\phi_2})=(0.45,0.40)$. All the other parameters are the same as in Fig.~\ref{Fig1:model}.
    }
    \label{Fig4:hyesteresis}
\end{figure*}

The phase diagram is qualitatively %
consistent with experiments in GUVs~\cite{cornell_tuning_2018}. In particular, when the mean composition is varied along any of the tie lines (a typical tie line is indicated by the black dashed line in Fig.~\ref{Fig1:model}B), the system always morphs from dots to stripes and eventually back to dots, which is the same as the sequence of morphologies observed in experiments~\cite{cornell_tuning_2018}. 
When varied in the normal (perpendicular) direction (indicated by the white dashed line in Fig.~\ref{Fig1:model}B), the patterns can switch between dots and stripes, as observed in experiments~\cite{cornell_tuning_2018}. The exact sequence of patterns is not unique: it depends on where the normal line crosses the tie line. 
The miscibility boundary observed here should be different from those measured for vesicles with no excess area~\cite{cornell_tuning_2018}. Hence, further experimental characterization of the phase diagram of patterns formed after introducing extra lipids will be informative for determining the model parameters and understanding the underlying interactions.

The morphology near the critical points has been suggested as an important criteria for distinguishing different mechanisms for the small domains~\cite{cornell_tuning_2018}. In particular, the observation of stripe domains near the critical point was used to rule out mechanisms involving bilayer curvature~\cite{cornell_tuning_2018, harden_budding_2005}. Here, however, the system can exhibit any of the morphologies near the two critical points (indicated by red stars), depending on the direction in which the critical point is approached.
Since all the phase boundaries become tangent to the miscibility boundary at the critical points, the stripe phase becomes dominant in its vicinity. In other words, if we draw a circle near the critical point $\bs{\phi}_c$ with radius $\epsilon$, both the stripe phase and the uniform phase will occupy half the area in the circle as $\epsilon\to 0$, with all the other morphologies occupying infinitesimal area. This may explain the experimental observation of stripes near the critical point.
It will be interesting to test experimentally whether departing the critical point in different directions eventually leads to different morphologies.

\textbf{Droplets coarsen in taut membranes.}~~The small domains discussed so far only emerge at relatively low surface tension. As the membrane becomes more taut, its behavior approaches a ternary mixture with either a uniform state or large domains. The transition to the uniform state can be described by a saddle-node bifurcation in the amplitude space, while the transition to the large domain phase is best understood as a crossover of free energy.
Fig.~\ref{Fig3:coarsening} illustrates these two types of transitions. 
For transition to the uniform state, the morphology parameter $g$ decreases with $\sigma$ (in the regime where $L_c$ stays finite) and eventually crosses the threshold $g=-1/15$ (Fig.~\ref{Fig3:coarsening}A), below which only the uniform state is stable. Throughout the transition, the deformation field $h(\bsr)$ can be well described by the sum of a few Fourier modes (Eq.~\eqref{Eq:amplitude expansion} and Fig.~\ref{Fig3:coarsening}B), but its amplitude vanishes at large $\sigma$ due to a saddle-node bifurcation. 
On the other hand, the transition to large liquid domains can be understood by a free energy crossover. For these large domains, the membrane remains flat ($h=\mathrm{const}.$) within bulk phases, which can no longer be described by the sum of Fourier modes (Fig.~\ref{Fig3:coarsening}D). 
The composition of the bulk phases can be determined by a convex hull construction with an extra free energy penalty $\frac{1}{2}\kappa c_1^2\phi_1^2$. These flat droplets always coarsen to reach the system size. 
As shown in Fig.~\ref{Fig3:coarsening}C, the free energy density of the patterns (orange dots) increases with $\sigma$ and eventually exceeds that of the flat droplet phase (red line). The crossover of free energy density defines a transition point beyond which patterns of small domains are no longer observed.

\textbf{Hysteresis of pattern morphologies.}~~The bistability of stripes and dots in the mixed state suggests the possibility for the hysteresis of pattern morphology due to cycling control parameters. Namely, it is possible to observe either or a mixture of these two morphologies depending on the history of the control parameter.
One of the most accessible parameters is membrane tension, which can be tuned by exterior osmotic pressure. Fig.~\ref{Fig4:hyesteresis} shows a hypothetical experiment where the external salt concentration is first increased and then decreased. Increasing salt concentration decreases the surface tension, which leads to a transition from dots to stripes; decreasing concentration leads to the opposite transition. Indeed, the backward transition takes place around $\sigma=0.185$, which is much larger than the threshold for the forward transition $\sigma=0.135$. 
Hysteresis is also observed when cycling temperature (see SI, section III), although we had to again assume that temperature dependence predominantly enters through the entropy term. 
Since the morphology control parameter [Eq.~\eqref{Eq:g definition}] is non-monotonic in $\sigma$, it may also be possible to observe hysteresis in stripe-dot-stripe transitions by tuning the exterior salt concentration.
These hysteresis experiments will help elucidate the validity of the bistable solutions found here, and the transition thresholds may be used to quantitatively determine parameters in our model. 

\textbf{Discussion.}~~In this paper, we have proposed a theoretical model to describe the patterns formed by small domains in lipid membranes. 
By coupling the bilayer curvature to the local lipid composition, the model successfully captures existing measurements of the pattern morphology and length scales.
Several future directions are proposed to further examine the validity of this mechanism. 
One possibility is to tune the membrane tension in a larger dynamical range and in cycles to search for a non-monotonic change in the domain size as well as hysteresis of pattern morphologies. 
Another direction is to examine the rich structures of the morphological phase diagram (Fig.~\ref{Fig1:model}B). In particular, the theory predicts that different patterns can be observed when approaching the critical point from different directions, which has not been explored experimentally.
It will also be interesting to see how the patterns depend on the amount of DPPC introduced to the outer leaflet, which affects the $c_1$ parameter in the model.
New experimental information will also enable further theoretical work, such as fitting the phase boundaries and inverse design of pattern morphologies~\cite{goodrich_designing_2021}.  

The patterns discussed here arise from equilibrium interactions governed by an overall free energy. 
Nonetheless, biological membranes under physiological conditions are highly nonequilibrium, influenced by various active processes that physically deform~\cite{stachowiak_costbenefit_2013} or chemically modify~\cite{falkenburger_symposium_2010,hsieh_dynamic_2021} the membrane. 
These nonequilibrium interactions enable rich spatiotemporal behaviors such as traveling waves of lipids~\cite{hsieh_dynamic_2021}, proteins~\cite{tan_topological_2020}, and membrane shape changes~\cite{wigbers_hierarchy_2021}.
By explicitly modeling the kinetics of nonequilibrium processes (e.g. phosphorylation/dephosphorylation cycles), the theoretical framework developed here can be extended to offer more insights into the morphology and dynamics of these nonequilibrium patterns.

\begin{acknowledgments}
    The work is supported by the National Science Foundation, through the Princeton Center for Complex Materials (DMR-2011750), and by the Princeton Catalysis Initiative.
    The authors thank S.~L.~Keller and I.~Levin for introducing us to this experimental system and for many stimulating discussions at various stages of this work. 
\end{acknowledgments}

\bibliography{membrane}

\begin{thebibliography}{47}%
\makeatletter
\providecommand \@ifxundefined [1]{%
 \@ifx{#1\undefined}
}%
\providecommand \@ifnum [1]{%
 \ifnum #1\expandafter \@firstoftwo
 \else \expandafter \@secondoftwo
 \fi
}%
\providecommand \@ifx [1]{%
 \ifx #1\expandafter \@firstoftwo
 \else \expandafter \@secondoftwo
 \fi
}%
\providecommand \natexlab [1]{#1}%
\providecommand \enquote  [1]{``#1''}%
\providecommand \bibnamefont  [1]{#1}%
\providecommand \bibfnamefont [1]{#1}%
\providecommand \citenamefont [1]{#1}%
\providecommand \href@noop [0]{\@secondoftwo}%
\providecommand \href [0]{\begingroup \@sanitize@url \@href}%
\providecommand \@href[1]{\@@startlink{#1}\@@href}%
\providecommand \@@href[1]{\endgroup#1\@@endlink}%
\providecommand \@sanitize@url [0]{\catcode `\\12\catcode `\$12\catcode
  `\&12\catcode `\#12\catcode `\^12\catcode `\_12\catcode `\%12\relax}%
\providecommand \@@startlink[1]{}%
\providecommand \@@endlink[0]{}%
\providecommand \url  [0]{\begingroup\@sanitize@url \@url }%
\providecommand \@url [1]{\endgroup\@href {#1}{\urlprefix }}%
\providecommand \urlprefix  [0]{URL }%
\providecommand \Eprint [0]{\href }%
\providecommand \doibase [0]{https://doi.org/}%
\providecommand \selectlanguage [0]{\@gobble}%
\providecommand \bibinfo  [0]{\@secondoftwo}%
\providecommand \bibfield  [0]{\@secondoftwo}%
\providecommand \translation [1]{[#1]}%
\providecommand \BibitemOpen [0]{}%
\providecommand \bibitemStop [0]{}%
\providecommand \bibitemNoStop [0]{.\EOS\space}%
\providecommand \EOS [0]{\spacefactor3000\relax}%
\providecommand \BibitemShut  [1]{\csname bibitem#1\endcsname}%
\let\auto@bib@innerbib\@empty
\bibitem [{\citenamefont {Semrau}\ and\ \citenamefont
  {Schmidt}(2009)}]{semrau_membrane_2009}%
  \BibitemOpen
  \bibfield  {author} {\bibinfo {author} {\bibfnamefont {S.}~\bibnamefont
  {Semrau}}\ and\ \bibinfo {author} {\bibfnamefont {T.}~\bibnamefont
  {Schmidt}},\ }\bibfield  {title} {\bibinfo {title} {Membrane heterogeneity
  – from lipid domains to curvature effects},\ }\href
  {https://doi.org/10.1039/b901587f} {\bibfield  {journal} {\bibinfo  {journal}
  {Soft Matter}\ }\textbf {\bibinfo {volume} {5}},\ \bibinfo {pages} {3174}
  (\bibinfo {year} {2009})}\BibitemShut {NoStop}%
\bibitem [{\citenamefont {Sych}\ \emph {et~al.}(2021)\citenamefont {Sych},
  \citenamefont {Gurdap}, \citenamefont {Wedemann},\ and\ \citenamefont
  {Sezgin}}]{sych_how_2021}%
  \BibitemOpen
  \bibfield  {author} {\bibinfo {author} {\bibfnamefont {T.}~\bibnamefont
  {Sych}}, \bibinfo {author} {\bibfnamefont {C.~O.}\ \bibnamefont {Gurdap}},
  \bibinfo {author} {\bibfnamefont {L.}~\bibnamefont {Wedemann}},\ and\
  \bibinfo {author} {\bibfnamefont {E.}~\bibnamefont {Sezgin}},\ }\bibfield
  {title} {\bibinfo {title} {How {Does} {Liquid}-{Liquid} {Phase} {Separation}
  in {Model} {Membranes} {Reflect} {Cell} {Membrane} {Heterogeneity}?},\ }\href
  {https://doi.org/10.3390/membranes11050323} {\bibfield  {journal} {\bibinfo
  {journal} {Membranes}\ }\textbf {\bibinfo {volume} {11}},\ \bibinfo {pages}
  {323} (\bibinfo {year} {2021})}\BibitemShut {NoStop}%
\bibitem [{\citenamefont {Baumgart}\ \emph {et~al.}(2007)\citenamefont
  {Baumgart}, \citenamefont {Hammond}, \citenamefont {Sengupta}, \citenamefont
  {Hess}, \citenamefont {Holowka}, \citenamefont {Baird},\ and\ \citenamefont
  {Webb}}]{baumgart_large-scale_2007}%
  \BibitemOpen
  \bibfield  {author} {\bibinfo {author} {\bibfnamefont {T.}~\bibnamefont
  {Baumgart}}, \bibinfo {author} {\bibfnamefont {A.~T.}\ \bibnamefont
  {Hammond}}, \bibinfo {author} {\bibfnamefont {P.}~\bibnamefont {Sengupta}},
  \bibinfo {author} {\bibfnamefont {S.~T.}\ \bibnamefont {Hess}}, \bibinfo
  {author} {\bibfnamefont {D.~A.}\ \bibnamefont {Holowka}}, \bibinfo {author}
  {\bibfnamefont {B.~A.}\ \bibnamefont {Baird}},\ and\ \bibinfo {author}
  {\bibfnamefont {W.~W.}\ \bibnamefont {Webb}},\ }\bibfield  {title} {\bibinfo
  {title} {Large-scale fluid/fluid phase separation of proteins and lipids in
  giant plasma membrane vesicles},\ }\href
  {https://doi.org/10.1073/pnas.0611357104} {\bibfield  {journal} {\bibinfo
  {journal} {Proc. Natl. Acad. Sci. U.S.A.}\ }\textbf {\bibinfo {volume}
  {104}},\ \bibinfo {pages} {3165} (\bibinfo {year} {2007})}\BibitemShut
  {NoStop}%
\bibitem [{\citenamefont {Dietrich}\ \emph {et~al.}(2001)\citenamefont
  {Dietrich}, \citenamefont {Bagatolli}, \citenamefont {Volovyk}, \citenamefont
  {Thompson}, \citenamefont {Levi}, \citenamefont {Jacobson},\ and\
  \citenamefont {Gratton}}]{dietrich_lipid_2001}%
  \BibitemOpen
  \bibfield  {author} {\bibinfo {author} {\bibfnamefont {C.}~\bibnamefont
  {Dietrich}}, \bibinfo {author} {\bibfnamefont {L.}~\bibnamefont {Bagatolli}},
  \bibinfo {author} {\bibfnamefont {Z.}~\bibnamefont {Volovyk}}, \bibinfo
  {author} {\bibfnamefont {N.}~\bibnamefont {Thompson}}, \bibinfo {author}
  {\bibfnamefont {M.}~\bibnamefont {Levi}}, \bibinfo {author} {\bibfnamefont
  {K.}~\bibnamefont {Jacobson}},\ and\ \bibinfo {author} {\bibfnamefont
  {E.}~\bibnamefont {Gratton}},\ }\bibfield  {title} {\bibinfo {title} {Lipid
  {Rafts} {Reconstituted} in {Model} {Membranes}},\ }\href
  {https://doi.org/10.1016/S0006-3495(01)76114-0} {\bibfield  {journal}
  {\bibinfo  {journal} {Biophysical Journal}\ }\textbf {\bibinfo {volume}
  {80}},\ \bibinfo {pages} {1417} (\bibinfo {year} {2001})}\BibitemShut
  {NoStop}%
\bibitem [{\citenamefont {Dimova}\ \emph {et~al.}(2006)\citenamefont {Dimova},
  \citenamefont {Aranda}, \citenamefont {Bezlyepkina}, \citenamefont {Nikolov},
  \citenamefont {Riske},\ and\ \citenamefont
  {Lipowsky}}]{dimova_practical_2006}%
  \BibitemOpen
  \bibfield  {author} {\bibinfo {author} {\bibfnamefont {R.}~\bibnamefont
  {Dimova}}, \bibinfo {author} {\bibfnamefont {S.}~\bibnamefont {Aranda}},
  \bibinfo {author} {\bibfnamefont {N.}~\bibnamefont {Bezlyepkina}}, \bibinfo
  {author} {\bibfnamefont {V.}~\bibnamefont {Nikolov}}, \bibinfo {author}
  {\bibfnamefont {K.~A.}\ \bibnamefont {Riske}},\ and\ \bibinfo {author}
  {\bibfnamefont {R.}~\bibnamefont {Lipowsky}},\ }\bibfield  {title} {\bibinfo
  {title} {A practical guide to giant vesicles. {Probing} the membrane
  nanoregime via optical microscopy},\ }\href
  {https://doi.org/10.1088/0953-8984/18/28/S04} {\bibfield  {journal} {\bibinfo
   {journal} {J. Phys.: Condens. Matter}\ }\textbf {\bibinfo {volume} {18}},\
  \bibinfo {pages} {S1151} (\bibinfo {year} {2006})}\BibitemShut {NoStop}%
\bibitem [{\citenamefont {Sengupta}\ \emph {et~al.}(2007)\citenamefont
  {Sengupta}, \citenamefont {Baird},\ and\ \citenamefont
  {Holowka}}]{sengupta_lipid_2007}%
  \BibitemOpen
  \bibfield  {author} {\bibinfo {author} {\bibfnamefont {P.}~\bibnamefont
  {Sengupta}}, \bibinfo {author} {\bibfnamefont {B.}~\bibnamefont {Baird}},\
  and\ \bibinfo {author} {\bibfnamefont {D.}~\bibnamefont {Holowka}},\
  }\bibfield  {title} {\bibinfo {title} {Lipid rafts, fluid/fluid phase
  separation, and their relevance to plasma membrane structure and function},\
  }\href {https://doi.org/10.1016/j.semcdb.2007.07.010} {\bibfield  {journal}
  {\bibinfo  {journal} {Seminars in Cell \& Developmental Biology}\ }\textbf
  {\bibinfo {volume} {18}},\ \bibinfo {pages} {583} (\bibinfo {year}
  {2007})}\BibitemShut {NoStop}%
\bibitem [{\citenamefont {Kabouridis}(2006)}]{kabouridis_lipid_2006}%
  \BibitemOpen
  \bibfield  {author} {\bibinfo {author} {\bibfnamefont {P.~S.}\ \bibnamefont
  {Kabouridis}},\ }\bibfield  {title} {\bibinfo {title} {Lipid rafts in {T}
  cell receptor signalling ({Review})},\ }\href
  {https://doi.org/10.1080/09687860500453673} {\bibfield  {journal} {\bibinfo
  {journal} {Molecular Membrane Biology}\ }\textbf {\bibinfo {volume} {23}},\
  \bibinfo {pages} {49} (\bibinfo {year} {2006})}\BibitemShut {NoStop}%
\bibitem [{\citenamefont {Mañes}\ and\ \citenamefont
  {Viola}(2006)}]{manes_lipid_2006}%
  \BibitemOpen
  \bibfield  {author} {\bibinfo {author} {\bibfnamefont {S.}~\bibnamefont
  {Mañes}}\ and\ \bibinfo {author} {\bibfnamefont {A.}~\bibnamefont {Viola}},\
  }\bibfield  {title} {\bibinfo {title} {Lipid rafts in lymphocyte activation
  and migration ({Review})},\ }\href
  {https://doi.org/10.1080/09687860500430069} {\bibfield  {journal} {\bibinfo
  {journal} {Molecular Membrane Biology}\ }\textbf {\bibinfo {volume} {23}},\
  \bibinfo {pages} {59} (\bibinfo {year} {2006})}\BibitemShut {NoStop}%
\bibitem [{\citenamefont {Parton}\ and\ \citenamefont
  {Richards}(2003)}]{parton_lipid_2003}%
  \BibitemOpen
  \bibfield  {author} {\bibinfo {author} {\bibfnamefont {R.~G.}\ \bibnamefont
  {Parton}}\ and\ \bibinfo {author} {\bibfnamefont {A.~A.}\ \bibnamefont
  {Richards}},\ }\bibfield  {title} {\bibinfo {title} {Lipid {Rafts} and
  {Caveolae} as {Portals} for {Endocytosis}: {New} {Insights} and {Common}
  {Mechanisms}: {{Caveolae}/{Lipid} {Raft}-{Dependent} {Endocytosis}}},\ }\href
  {https://doi.org/10.1034/j.1600-0854.2003.00128.x} {\bibfield  {journal}
  {\bibinfo  {journal} {Traffic}\ }\textbf {\bibinfo {volume} {4}},\ \bibinfo
  {pages} {724} (\bibinfo {year} {2003})}\BibitemShut {NoStop}%
\bibitem [{\citenamefont {Salaün}\ \emph {et~al.}(2004)\citenamefont
  {Salaün}, \citenamefont {James},\ and\ \citenamefont
  {Chamberlain}}]{salaun_lipid_2004}%
  \BibitemOpen
  \bibfield  {author} {\bibinfo {author} {\bibfnamefont {C.}~\bibnamefont
  {Salaün}}, \bibinfo {author} {\bibfnamefont {D.~J.}\ \bibnamefont {James}},\
  and\ \bibinfo {author} {\bibfnamefont {L.~H.}\ \bibnamefont {Chamberlain}},\
  }\bibfield  {title} {\bibinfo {title} {Lipid {Rafts} and the {Regulation} of
  {Exocytosis}: {Lipid} {Rafts} and {Exocytosis}},\ }\href
  {https://doi.org/10.1111/j.1600-0854.2004.0162.x} {\bibfield  {journal}
  {\bibinfo  {journal} {Traffic}\ }\textbf {\bibinfo {volume} {5}},\ \bibinfo
  {pages} {255} (\bibinfo {year} {2004})}\BibitemShut {NoStop}%
\bibitem [{\citenamefont {Murley}\ \emph {et~al.}(2017)\citenamefont {Murley},
  \citenamefont {Yamada}, \citenamefont {Niles}, \citenamefont {Toulmay},
  \citenamefont {Prinz}, \citenamefont {Powers},\ and\ \citenamefont
  {Nunnari}}]{murley_sterol_2017}%
  \BibitemOpen
  \bibfield  {author} {\bibinfo {author} {\bibfnamefont {A.}~\bibnamefont
  {Murley}}, \bibinfo {author} {\bibfnamefont {J.}~\bibnamefont {Yamada}},
  \bibinfo {author} {\bibfnamefont {B.~J.}\ \bibnamefont {Niles}}, \bibinfo
  {author} {\bibfnamefont {A.}~\bibnamefont {Toulmay}}, \bibinfo {author}
  {\bibfnamefont {W.~A.}\ \bibnamefont {Prinz}}, \bibinfo {author}
  {\bibfnamefont {T.}~\bibnamefont {Powers}},\ and\ \bibinfo {author}
  {\bibfnamefont {J.}~\bibnamefont {Nunnari}},\ }\bibfield  {title} {\bibinfo
  {title} {Sterol transporters at membrane contact sites regulate {TORC1} and
  {TORC2} signaling},\ }\href {https://doi.org/10.1083/jcb.201610032}
  {\bibfield  {journal} {\bibinfo  {journal} {Journal of Cell Biology}\
  }\textbf {\bibinfo {volume} {216}},\ \bibinfo {pages} {2679} (\bibinfo {year}
  {2017})}\BibitemShut {NoStop}%
\bibitem [{\citenamefont {Seo}\ \emph {et~al.}(2017)\citenamefont {Seo},
  \citenamefont {Lau}, \citenamefont {Feliciano}, \citenamefont {Sengupta},
  \citenamefont {Gros}, \citenamefont {Cinquin}, \citenamefont {Larabell},\
  and\ \citenamefont {Lippincott-Schwartz}}]{seo_ampk_2017}%
  \BibitemOpen
  \bibfield  {author} {\bibinfo {author} {\bibfnamefont {A.~Y.}\ \bibnamefont
  {Seo}}, \bibinfo {author} {\bibfnamefont {P.-W.}\ \bibnamefont {Lau}},
  \bibinfo {author} {\bibfnamefont {D.}~\bibnamefont {Feliciano}}, \bibinfo
  {author} {\bibfnamefont {P.}~\bibnamefont {Sengupta}}, \bibinfo {author}
  {\bibfnamefont {M.~A.~L.}\ \bibnamefont {Gros}}, \bibinfo {author}
  {\bibfnamefont {B.}~\bibnamefont {Cinquin}}, \bibinfo {author} {\bibfnamefont
  {C.~A.}\ \bibnamefont {Larabell}},\ and\ \bibinfo {author} {\bibfnamefont
  {J.}~\bibnamefont {Lippincott-Schwartz}},\ }\bibfield  {title} {\bibinfo
  {title} {{AMPK} and vacuole-associated {Atg14p} orchestrate mu-lipophagy for
  energy production and long-term survival under glucose starvation},\ }\href
  {https://doi.org/10.7554/eLife.21690} {\bibfield  {journal} {\bibinfo
  {journal} {eLife}\ }\textbf {\bibinfo {volume} {6}},\ \bibinfo {pages}
  {e21690} (\bibinfo {year} {2017})}\BibitemShut {NoStop}%
\bibitem [{\citenamefont {Leveille}\ \emph {et~al.}(2022)\citenamefont
  {Leveille}, \citenamefont {Cornell}, \citenamefont {Merz},\ and\
  \citenamefont {Keller}}]{leveille_yeast_2022}%
  \BibitemOpen
  \bibfield  {author} {\bibinfo {author} {\bibfnamefont {C.~L.}\ \bibnamefont
  {Leveille}}, \bibinfo {author} {\bibfnamefont {C.~E.}\ \bibnamefont
  {Cornell}}, \bibinfo {author} {\bibfnamefont {A.~J.}\ \bibnamefont {Merz}},\
  and\ \bibinfo {author} {\bibfnamefont {S.~L.}\ \bibnamefont {Keller}},\
  }\bibfield  {title} {\bibinfo {title} {Yeast cells actively tune their
  membranes to phase separate at temperatures that scale with growth
  temperatures},\ }\href {https://doi.org/10.1073/pnas.2116007119} {\bibfield
  {journal} {\bibinfo  {journal} {Proc. Natl. Acad. Sci. U.S.A.}\ }\textbf
  {\bibinfo {volume} {119}},\ \bibinfo {pages} {e2116007119} (\bibinfo {year}
  {2022})}\BibitemShut {NoStop}%
\bibitem [{\citenamefont {Veatch}\ and\ \citenamefont
  {Keller}(2002)}]{veatch_organization_2002}%
  \BibitemOpen
  \bibfield  {author} {\bibinfo {author} {\bibfnamefont {S.~L.}\ \bibnamefont
  {Veatch}}\ and\ \bibinfo {author} {\bibfnamefont {S.~L.}\ \bibnamefont
  {Keller}},\ }\bibfield  {title} {\bibinfo {title} {Organization in {Lipid}
  {Membranes} {Containing} {Cholesterol}},\ }\href
  {https://doi.org/10.1103/PhysRevLett.89.268101} {\bibfield  {journal}
  {\bibinfo  {journal} {Phys. Rev. Lett.}\ }\textbf {\bibinfo {volume} {89}},\
  \bibinfo {pages} {268101} (\bibinfo {year} {2002})}\BibitemShut {NoStop}%
\bibitem [{\citenamefont {Veatch}\ and\ \citenamefont
  {Keller}(2003)}]{veatch_separation_2003}%
  \BibitemOpen
  \bibfield  {author} {\bibinfo {author} {\bibfnamefont {S.~L.}\ \bibnamefont
  {Veatch}}\ and\ \bibinfo {author} {\bibfnamefont {S.~L.}\ \bibnamefont
  {Keller}},\ }\bibfield  {title} {\bibinfo {title} {Separation of {Liquid}
  {Phases} in {Giant} {Vesicles} of {Ternary} {Mixtures} of {Phospholipids} and
  {Cholesterol}},\ }\href {https://doi.org/10.1016/S0006-3495(03)74726-2}
  {\bibfield  {journal} {\bibinfo  {journal} {Biophysical Journal}\ }\textbf
  {\bibinfo {volume} {85}},\ \bibinfo {pages} {3074} (\bibinfo {year}
  {2003})}\BibitemShut {NoStop}%
\bibitem [{\citenamefont {Stanich}\ \emph {et~al.}(2013)\citenamefont
  {Stanich}, \citenamefont {Honerkamp-Smith}, \citenamefont {Putzel},
  \citenamefont {Warth}, \citenamefont {Lamprecht}, \citenamefont {Mandal},
  \citenamefont {Mann}, \citenamefont {Hua},\ and\ \citenamefont
  {Keller}}]{stanich_coarsening_2013}%
  \BibitemOpen
  \bibfield  {author} {\bibinfo {author} {\bibfnamefont {C.}~\bibnamefont
  {Stanich}}, \bibinfo {author} {\bibfnamefont {A.}~\bibnamefont
  {Honerkamp-Smith}}, \bibinfo {author} {\bibfnamefont {G.}~\bibnamefont
  {Putzel}}, \bibinfo {author} {\bibfnamefont {C.}~\bibnamefont {Warth}},
  \bibinfo {author} {\bibfnamefont {A.}~\bibnamefont {Lamprecht}}, \bibinfo
  {author} {\bibfnamefont {P.}~\bibnamefont {Mandal}}, \bibinfo {author}
  {\bibfnamefont {E.}~\bibnamefont {Mann}}, \bibinfo {author} {\bibfnamefont
  {T.-A.}\ \bibnamefont {Hua}},\ and\ \bibinfo {author} {\bibfnamefont
  {S.}~\bibnamefont {Keller}},\ }\bibfield  {title} {\bibinfo {title}
  {Coarsening {Dynamics} of {Domains} in {Lipid} {Membranes}},\ }\href
  {https://doi.org/10.1016/j.bpj.2013.06.013} {\bibfield  {journal} {\bibinfo
  {journal} {Biophysical Journal}\ }\textbf {\bibinfo {volume} {105}},\
  \bibinfo {pages} {444} (\bibinfo {year} {2013})}\BibitemShut {NoStop}%
\bibitem [{\citenamefont {Cornell}\ \emph {et~al.}(2018)\citenamefont
  {Cornell}, \citenamefont {Skinkle}, \citenamefont {He}, \citenamefont
  {Levental}, \citenamefont {Levental},\ and\ \citenamefont
  {Keller}}]{cornell_tuning_2018}%
  \BibitemOpen
  \bibfield  {author} {\bibinfo {author} {\bibfnamefont {C.~E.}\ \bibnamefont
  {Cornell}}, \bibinfo {author} {\bibfnamefont {A.~D.}\ \bibnamefont
  {Skinkle}}, \bibinfo {author} {\bibfnamefont {S.}~\bibnamefont {He}},
  \bibinfo {author} {\bibfnamefont {I.}~\bibnamefont {Levental}}, \bibinfo
  {author} {\bibfnamefont {K.~R.}\ \bibnamefont {Levental}},\ and\ \bibinfo
  {author} {\bibfnamefont {S.~L.}\ \bibnamefont {Keller}},\ }\bibfield  {title}
  {\bibinfo {title} {Tuning {Length} {Scales} of {Small} {Domains} in
  {Cell}-{Derived} {Membranes} and {Synthetic} {Model} {Membranes}},\ }\href
  {https://doi.org/10.1016/j.bpj.2018.06.027} {\bibfield  {journal} {\bibinfo
  {journal} {Biophysical Journal}\ }\textbf {\bibinfo {volume} {115}},\
  \bibinfo {pages} {690} (\bibinfo {year} {2018})}\BibitemShut {NoStop}%
\bibitem [{\citenamefont {Schmid}(2017)}]{schmid_physical_2017}%
  \BibitemOpen
  \bibfield  {author} {\bibinfo {author} {\bibfnamefont {F.}~\bibnamefont
  {Schmid}},\ }\bibfield  {title} {\bibinfo {title} {Physical mechanisms of
  micro- and nanodomain formation in multicomponent lipid membranes},\ }\href
  {https://doi.org/10.1016/j.bbamem.2016.10.021} {\bibfield  {journal}
  {\bibinfo  {journal} {Biochimica et Biophysica Acta (BBA) - Biomembranes}\
  }\textbf {\bibinfo {volume} {1859}},\ \bibinfo {pages} {509} (\bibinfo {year}
  {2017})}\BibitemShut {NoStop}%
\bibitem [{\citenamefont {Schick}(2018)}]{schick_strongly_2018}%
  \BibitemOpen
  \bibfield  {author} {\bibinfo {author} {\bibfnamefont {M.}~\bibnamefont
  {Schick}},\ }\bibfield  {title} {\bibinfo {title} {Strongly {Correlated}
  {Rafts} in {Both} {Leaves} of an {Asymmetric} {Bilayer}},\ }\href
  {https://doi.org/10.1021/acs.jpcb.7b08890} {\bibfield  {journal} {\bibinfo
  {journal} {J. Phys. Chem. B}\ }\textbf {\bibinfo {volume} {122}},\ \bibinfo
  {pages} {3251} (\bibinfo {year} {2018})}\BibitemShut {NoStop}%
\bibitem [{\citenamefont {Allender}\ and\ \citenamefont
  {Schick}(2022)}]{allender_theoretical_2022}%
  \BibitemOpen
  \bibfield  {author} {\bibinfo {author} {\bibfnamefont {D.~W.}\ \bibnamefont
  {Allender}}\ and\ \bibinfo {author} {\bibfnamefont {M.}~\bibnamefont
  {Schick}},\ }\bibfield  {title} {\bibinfo {title} {A {Theoretical} {Basis}
  for {Nanodomains}},\ }\href {https://doi.org/10.1007/s00232-021-00213-x}
  {\bibfield  {journal} {\bibinfo  {journal} {J Membrane Biol}\ }\textbf
  {\bibinfo {volume} {255}},\ \bibinfo {pages} {451} (\bibinfo {year}
  {2022})}\BibitemShut {NoStop}%
\bibitem [{\citenamefont {Harden}\ \emph {et~al.}(2005)\citenamefont {Harden},
  \citenamefont {MacKintosh},\ and\ \citenamefont
  {Olmsted}}]{harden_budding_2005}%
  \BibitemOpen
  \bibfield  {author} {\bibinfo {author} {\bibfnamefont {J.}~\bibnamefont
  {Harden}}, \bibinfo {author} {\bibfnamefont {F.}~\bibnamefont {MacKintosh}},\
  and\ \bibinfo {author} {\bibfnamefont {P.}~\bibnamefont {Olmsted}},\
  }\bibfield  {title} {\bibinfo {title} {Budding and domain shape
  transformations in mixed lipid films and bilayer membranes},\ }\href
  {https://doi.org/10.1103/PhysRevE.72.011903} {\bibfield  {journal} {\bibinfo
  {journal} {Phys. Rev. E}\ }\textbf {\bibinfo {volume} {72}},\ \bibinfo
  {pages} {011903} (\bibinfo {year} {2005})}\BibitemShut {NoStop}%
\bibitem [{\citenamefont {Usery}\ \emph {et~al.}(2017)\citenamefont {Usery},
  \citenamefont {Enoki}, \citenamefont {Wickramasinghe}, \citenamefont
  {Weiner}, \citenamefont {Tsai}, \citenamefont {Kim}, \citenamefont {Wang},
  \citenamefont {Torng}, \citenamefont {Ackerman}, \citenamefont {Heberle},
  \citenamefont {Katsaras},\ and\ \citenamefont {Feigenson}}]{usery_line_2017}%
  \BibitemOpen
  \bibfield  {author} {\bibinfo {author} {\bibfnamefont {R.~D.}\ \bibnamefont
  {Usery}}, \bibinfo {author} {\bibfnamefont {T.~A.}\ \bibnamefont {Enoki}},
  \bibinfo {author} {\bibfnamefont {S.~P.}\ \bibnamefont {Wickramasinghe}},
  \bibinfo {author} {\bibfnamefont {M.~D.}\ \bibnamefont {Weiner}}, \bibinfo
  {author} {\bibfnamefont {W.-C.}\ \bibnamefont {Tsai}}, \bibinfo {author}
  {\bibfnamefont {M.~B.}\ \bibnamefont {Kim}}, \bibinfo {author} {\bibfnamefont
  {S.}~\bibnamefont {Wang}}, \bibinfo {author} {\bibfnamefont {T.~L.}\
  \bibnamefont {Torng}}, \bibinfo {author} {\bibfnamefont {D.~G.}\ \bibnamefont
  {Ackerman}}, \bibinfo {author} {\bibfnamefont {F.~A.}\ \bibnamefont
  {Heberle}}, \bibinfo {author} {\bibfnamefont {J.}~\bibnamefont {Katsaras}},\
  and\ \bibinfo {author} {\bibfnamefont {G.~W.}\ \bibnamefont {Feigenson}},\
  }\bibfield  {title} {\bibinfo {title} {Line {Tension} {Controls}
  {Liquid}-{Disordered} + {Liquid}-{Ordered} {Domain} {Size} {Transition} in
  {Lipid} {Bilayers}},\ }\href {https://doi.org/10.1016/j.bpj.2017.02.033}
  {\bibfield  {journal} {\bibinfo  {journal} {Biophysical Journal}\ }\textbf
  {\bibinfo {volume} {112}},\ \bibinfo {pages} {1431} (\bibinfo {year}
  {2017})}\BibitemShut {NoStop}%
\bibitem [{\citenamefont {Honerkamp-Smith}\ \emph {et~al.}(2012)\citenamefont
  {Honerkamp-Smith}, \citenamefont {Machta},\ and\ \citenamefont
  {Keller}}]{honerkamp-smith_experimental_2012}%
  \BibitemOpen
  \bibfield  {author} {\bibinfo {author} {\bibfnamefont {A.~R.}\ \bibnamefont
  {Honerkamp-Smith}}, \bibinfo {author} {\bibfnamefont {B.~B.}\ \bibnamefont
  {Machta}},\ and\ \bibinfo {author} {\bibfnamefont {S.~L.}\ \bibnamefont
  {Keller}},\ }\bibfield  {title} {\bibinfo {title} {Experimental
  {Observations} of {Dynamic} {Critical} {Phenomena} in a {Lipid} {Membrane}},\
  }\href {https://doi.org/10.1103/PhysRevLett.108.265702} {\bibfield  {journal}
  {\bibinfo  {journal} {Phys. Rev. Lett.}\ }\textbf {\bibinfo {volume} {108}},\
  \bibinfo {pages} {265702} (\bibinfo {year} {2012})}\BibitemShut {NoStop}%
\bibitem [{\citenamefont {Hirose}\ \emph {et~al.}(2012)\citenamefont {Hirose},
  \citenamefont {Komura},\ and\ \citenamefont
  {Andelman}}]{hirose_concentration_2012}%
  \BibitemOpen
  \bibfield  {author} {\bibinfo {author} {\bibfnamefont {Y.}~\bibnamefont
  {Hirose}}, \bibinfo {author} {\bibfnamefont {S.}~\bibnamefont {Komura}},\
  and\ \bibinfo {author} {\bibfnamefont {D.}~\bibnamefont {Andelman}},\
  }\bibfield  {title} {\bibinfo {title} {Concentration fluctuations and phase
  transitions in coupled modulated bilayers},\ }\href
  {https://doi.org/10.1103/PhysRevE.86.021916} {\bibfield  {journal} {\bibinfo
  {journal} {Phys. Rev. E}\ }\textbf {\bibinfo {volume} {86}},\ \bibinfo
  {pages} {021916} (\bibinfo {year} {2012})}\BibitemShut {NoStop}%
\bibitem [{\citenamefont {Flory}(1942)}]{flory_thermodynamics_1942}%
  \BibitemOpen
  \bibfield  {author} {\bibinfo {author} {\bibfnamefont {P.~J.}\ \bibnamefont
  {Flory}},\ }\bibfield  {title} {\bibinfo {title} {Thermodynamics of {High}
  {Polymer} {Solutions}},\ }\href {https://doi.org/10.1063/1.1723621}
  {\bibfield  {journal} {\bibinfo  {journal} {The Journal of Chemical Physics}\
  }\textbf {\bibinfo {volume} {10}},\ \bibinfo {pages} {51} (\bibinfo {year}
  {1942})}\BibitemShut {NoStop}%
\bibitem [{\citenamefont {Huggins}(1941)}]{huggins_solutions_1941}%
  \BibitemOpen
  \bibfield  {author} {\bibinfo {author} {\bibfnamefont {M.~L.}\ \bibnamefont
  {Huggins}},\ }\bibfield  {title} {\bibinfo {title} {Solutions of {Long}
  {Chain} {Compounds}},\ }\href {https://doi.org/10.1063/1.1750930} {\bibfield
  {journal} {\bibinfo  {journal} {The Journal of Chemical Physics}\ }\textbf
  {\bibinfo {volume} {9}},\ \bibinfo {pages} {440} (\bibinfo {year}
  {1941})}\BibitemShut {NoStop}%
\bibitem [{\citenamefont {Idema}\ \emph {et~al.}(2009)\citenamefont {Idema},
  \citenamefont {van Leeuwen},\ and\ \citenamefont {Storm}}]{idema_phase_2009}%
  \BibitemOpen
  \bibfield  {author} {\bibinfo {author} {\bibfnamefont {T.}~\bibnamefont
  {Idema}}, \bibinfo {author} {\bibfnamefont {J.~M.~J.}\ \bibnamefont {van
  Leeuwen}},\ and\ \bibinfo {author} {\bibfnamefont {C.}~\bibnamefont
  {Storm}},\ }\bibfield  {title} {\bibinfo {title} {Phase coexistence and line
  tension in ternary lipid systems},\ }\href
  {https://doi.org/10.1103/PhysRevE.80.041924} {\bibfield  {journal} {\bibinfo
  {journal} {Phys. Rev. E}\ }\textbf {\bibinfo {volume} {80}},\ \bibinfo
  {pages} {041924} (\bibinfo {year} {2009})}\BibitemShut {NoStop}%
\bibitem [{\citenamefont {Veatch}\ \emph {et~al.}(2006)\citenamefont {Veatch},
  \citenamefont {Gawrisch},\ and\ \citenamefont
  {Keller}}]{veatch_closed-loop_2006}%
  \BibitemOpen
  \bibfield  {author} {\bibinfo {author} {\bibfnamefont {S.~L.}\ \bibnamefont
  {Veatch}}, \bibinfo {author} {\bibfnamefont {K.}~\bibnamefont {Gawrisch}},\
  and\ \bibinfo {author} {\bibfnamefont {S.~L.}\ \bibnamefont {Keller}},\
  }\bibfield  {title} {\bibinfo {title} {Closed-{Loop} {Miscibility} {Gap} and
  {Quantitative} {Tie}-{Lines} in {Ternary} {Membranes} {Containing}
  {Diphytanoyl} {PC}},\ }\href {https://doi.org/10.1529/biophysj.105.080283}
  {\bibfield  {journal} {\bibinfo  {journal} {Biophysical Journal}\ }\textbf
  {\bibinfo {volume} {90}},\ \bibinfo {pages} {4428} (\bibinfo {year}
  {2006})}\BibitemShut {NoStop}%
\bibitem [{\citenamefont {Hohenberg}\ and\ \citenamefont
  {Halperin}(1977)}]{hohenberg_theory_1977}%
  \BibitemOpen
  \bibfield  {author} {\bibinfo {author} {\bibfnamefont {P.~C.}\ \bibnamefont
  {Hohenberg}}\ and\ \bibinfo {author} {\bibfnamefont {B.~I.}\ \bibnamefont
  {Halperin}},\ }\bibfield  {title} {\bibinfo {title} {Theory of dynamic
  critical phenomena},\ }\href {https://doi.org/10.1103/RevModPhys.49.435}
  {\bibfield  {journal} {\bibinfo  {journal} {Rev. Mod. Phys.}\ }\textbf
  {\bibinfo {volume} {49}},\ \bibinfo {pages} {435} (\bibinfo {year}
  {1977})}\BibitemShut {NoStop}%
\bibitem [{\citenamefont {Yanagisawa}\ \emph {et~al.}(2007)\citenamefont
  {Yanagisawa}, \citenamefont {Imai}, \citenamefont {Masui}, \citenamefont
  {Komura},\ and\ \citenamefont {Ohta}}]{yanagisawa_growth_2007}%
  \BibitemOpen
  \bibfield  {author} {\bibinfo {author} {\bibfnamefont {M.}~\bibnamefont
  {Yanagisawa}}, \bibinfo {author} {\bibfnamefont {M.}~\bibnamefont {Imai}},
  \bibinfo {author} {\bibfnamefont {T.}~\bibnamefont {Masui}}, \bibinfo
  {author} {\bibfnamefont {S.}~\bibnamefont {Komura}},\ and\ \bibinfo {author}
  {\bibfnamefont {T.}~\bibnamefont {Ohta}},\ }\bibfield  {title} {\bibinfo
  {title} {Growth {Dynamics} of {Domains} in {Ternary} {Fluid} {Vesicles}},\
  }\href {https://doi.org/10.1529/biophysj.106.087494} {\bibfield  {journal}
  {\bibinfo  {journal} {Biophysical Journal}\ }\textbf {\bibinfo {volume}
  {92}},\ \bibinfo {pages} {115} (\bibinfo {year} {2007})}\BibitemShut
  {NoStop}%
\bibitem [{\citenamefont {Thiele}\ \emph {et~al.}(2013)\citenamefont {Thiele},
  \citenamefont {Archer}, \citenamefont {Robbins}, \citenamefont {Gomez},\ and\
  \citenamefont {Knobloch}}]{thiele_localized_2013}%
  \BibitemOpen
  \bibfield  {author} {\bibinfo {author} {\bibfnamefont {U.}~\bibnamefont
  {Thiele}}, \bibinfo {author} {\bibfnamefont {A.~J.}\ \bibnamefont {Archer}},
  \bibinfo {author} {\bibfnamefont {M.~J.}\ \bibnamefont {Robbins}}, \bibinfo
  {author} {\bibfnamefont {H.}~\bibnamefont {Gomez}},\ and\ \bibinfo {author}
  {\bibfnamefont {E.}~\bibnamefont {Knobloch}},\ }\bibfield  {title} {\bibinfo
  {title} {Localized states in the conserved {Swift}-{Hohenberg} equation with
  cubic nonlinearity},\ }\href@noop {} {\bibfield  {journal} {\bibinfo
  {journal} {PHYSICAL REVIEW E}\ ,\ \bibinfo {pages} {19}} (\bibinfo {year}
  {2013})}\BibitemShut {NoStop}%
\bibitem [{\citenamefont {Knobloch}(2016)}]{knobloch_localized_2016}%
  \BibitemOpen
  \bibfield  {author} {\bibinfo {author} {\bibfnamefont {E.}~\bibnamefont
  {Knobloch}},\ }\bibfield  {title} {\bibinfo {title} {Localized structures and
  front propagation in systems with a conservation law},\ }\href
  {https://doi.org/10.1093/imamat/hxw029} {\bibfield  {journal} {\bibinfo
  {journal} {IMA J Appl Math}\ }\textbf {\bibinfo {volume} {81}},\ \bibinfo
  {pages} {457} (\bibinfo {year} {2016})}\BibitemShut {NoStop}%
\bibitem [{\citenamefont {Emmerich}\ \emph {et~al.}(2012)\citenamefont
  {Emmerich}, \citenamefont {Löwen}, \citenamefont {Wittkowski}, \citenamefont
  {Gruhn}, \citenamefont {Tóth}, \citenamefont {Tegze},\ and\ \citenamefont
  {Gránásy}}]{emmerich_phase-field-crystal_2012}%
  \BibitemOpen
  \bibfield  {author} {\bibinfo {author} {\bibfnamefont {H.}~\bibnamefont
  {Emmerich}}, \bibinfo {author} {\bibfnamefont {H.}~\bibnamefont {Löwen}},
  \bibinfo {author} {\bibfnamefont {R.}~\bibnamefont {Wittkowski}}, \bibinfo
  {author} {\bibfnamefont {T.}~\bibnamefont {Gruhn}}, \bibinfo {author}
  {\bibfnamefont {G.~I.}\ \bibnamefont {Tóth}}, \bibinfo {author}
  {\bibfnamefont {G.}~\bibnamefont {Tegze}},\ and\ \bibinfo {author}
  {\bibfnamefont {L.}~\bibnamefont {Gránásy}},\ }\bibfield  {title} {\bibinfo
  {title} {Phase-field-crystal models for condensed matter dynamics on atomic
  length and diffusive time scales: an overview},\ }\href
  {https://doi.org/10.1080/00018732.2012.737555} {\bibfield  {journal}
  {\bibinfo  {journal} {Advances in Physics}\ }\textbf {\bibinfo {volume}
  {61}},\ \bibinfo {pages} {665} (\bibinfo {year} {2012})}\BibitemShut
  {NoStop}%
\bibitem [{\citenamefont {Elder}\ \emph {et~al.}(2010)\citenamefont {Elder},
  \citenamefont {Huang},\ and\ \citenamefont
  {Provatas}}]{elder_amplitude_2010}%
  \BibitemOpen
  \bibfield  {author} {\bibinfo {author} {\bibfnamefont {K.~R.}\ \bibnamefont
  {Elder}}, \bibinfo {author} {\bibfnamefont {Z.-F.}\ \bibnamefont {Huang}},\
  and\ \bibinfo {author} {\bibfnamefont {N.}~\bibnamefont {Provatas}},\
  }\bibfield  {title} {\bibinfo {title} {Amplitude expansion of the binary
  phase-field-crystal model},\ }\href
  {https://doi.org/10.1103/PhysRevE.81.011602} {\bibfield  {journal} {\bibinfo
  {journal} {Phys. Rev. E}\ }\textbf {\bibinfo {volume} {81}},\ \bibinfo
  {pages} {011602} (\bibinfo {year} {2010})}\BibitemShut {NoStop}%
\bibitem [{\citenamefont {Stefanovic}\ \emph {et~al.}(2006)\citenamefont
  {Stefanovic}, \citenamefont {Haataja},\ and\ \citenamefont
  {Provatas}}]{stefanovic_phase-field_2006}%
  \BibitemOpen
  \bibfield  {author} {\bibinfo {author} {\bibfnamefont {P.}~\bibnamefont
  {Stefanovic}}, \bibinfo {author} {\bibfnamefont {M.}~\bibnamefont
  {Haataja}},\ and\ \bibinfo {author} {\bibfnamefont {N.}~\bibnamefont
  {Provatas}},\ }\bibfield  {title} {\bibinfo {title} {Phase-{Field} {Crystals}
  with {Elastic} {Interactions}},\ }\href
  {https://doi.org/10.1103/PhysRevLett.96.225504} {\bibfield  {journal}
  {\bibinfo  {journal} {Phys. Rev. Lett.}\ }\textbf {\bibinfo {volume} {96}},\
  \bibinfo {pages} {225504} (\bibinfo {year} {2006})}\BibitemShut {NoStop}%
\bibitem [{\citenamefont {Matthews}\ and\ \citenamefont
  {Cox}(2000)}]{matthews_pattern_2000}%
  \BibitemOpen
  \bibfield  {author} {\bibinfo {author} {\bibfnamefont {P.~C.}\ \bibnamefont
  {Matthews}}\ and\ \bibinfo {author} {\bibfnamefont {S.~M.}\ \bibnamefont
  {Cox}},\ }\bibfield  {title} {\bibinfo {title} {Pattern formation with a
  conservation law},\ }\href {https://doi.org/10.1088/0951-7715/13/4/317}
  {\bibfield  {journal} {\bibinfo  {journal} {Nonlinearity}\ }\textbf {\bibinfo
  {volume} {13}},\ \bibinfo {pages} {1293} (\bibinfo {year}
  {2000})}\BibitemShut {NoStop}%
\bibitem [{\citenamefont {Cox}(2004)}]{cox_envelope_2004}%
  \BibitemOpen
  \bibfield  {author} {\bibinfo {author} {\bibfnamefont {S.~M.}\ \bibnamefont
  {Cox}},\ }\bibfield  {title} {\bibinfo {title} {The envelope of a
  one-dimensional pattern in the presence of a conserved quantity},\ }\href
  {https://doi.org/10.1016/j.physleta.2004.10.038} {\bibfield  {journal}
  {\bibinfo  {journal} {Physics Letters A}\ }\textbf {\bibinfo {volume}
  {333}},\ \bibinfo {pages} {91} (\bibinfo {year} {2004})}\BibitemShut
  {NoStop}%
\bibitem [{\citenamefont {Lifshitz}\ and\ \citenamefont
  {Slyozov}(1961)}]{lifshitz_kinetics_1961}%
  \BibitemOpen
  \bibfield  {author} {\bibinfo {author} {\bibfnamefont {I.}~\bibnamefont
  {Lifshitz}}\ and\ \bibinfo {author} {\bibfnamefont {V.}~\bibnamefont
  {Slyozov}},\ }\bibfield  {title} {\bibinfo {title} {The kinetics of
  precipitation from supersaturated solid solutions},\ }\href
  {https://doi.org/10.1016/0022-3697(61)90054-3} {\bibfield  {journal}
  {\bibinfo  {journal} {Journal of Physics and Chemistry of Solids}\ }\textbf
  {\bibinfo {volume} {19}},\ \bibinfo {pages} {35} (\bibinfo {year}
  {1961})}\BibitemShut {NoStop}%
\bibitem [{\citenamefont {Camley}\ and\ \citenamefont
  {Brown}(2011)}]{camley_dynamic_2011}%
  \BibitemOpen
  \bibfield  {author} {\bibinfo {author} {\bibfnamefont {B.~A.}\ \bibnamefont
  {Camley}}\ and\ \bibinfo {author} {\bibfnamefont {F.~L.~H.}\ \bibnamefont
  {Brown}},\ }\bibfield  {title} {\bibinfo {title} {Dynamic scaling in phase
  separation kinetics for quasi-two-dimensional membranes},\ }\href
  {https://doi.org/10.1063/1.3662131} {\bibfield  {journal} {\bibinfo
  {journal} {The Journal of Chemical Physics}\ }\textbf {\bibinfo {volume}
  {135}},\ \bibinfo {pages} {225106} (\bibinfo {year} {2011})}\BibitemShut
  {NoStop}%
\bibitem [{\citenamefont {Mao}\ \emph {et~al.}(2019)\citenamefont {Mao},
  \citenamefont {Kuldinow}, \citenamefont {Haataja},\ and\ \citenamefont
  {Košmrlj}}]{mao_phase_2019}%
  \BibitemOpen
  \bibfield  {author} {\bibinfo {author} {\bibfnamefont {S.}~\bibnamefont
  {Mao}}, \bibinfo {author} {\bibfnamefont {D.}~\bibnamefont {Kuldinow}},
  \bibinfo {author} {\bibfnamefont {M.~P.}\ \bibnamefont {Haataja}},\ and\
  \bibinfo {author} {\bibfnamefont {A.}~\bibnamefont {Košmrlj}},\ }\bibfield
  {title} {\bibinfo {title} {Phase behavior and morphology of multicomponent
  liquid mixtures},\ }\href {https://doi.org/10.1039/C8SM02045K} {\bibfield
  {journal} {\bibinfo  {journal} {Soft Matter}\ }\textbf {\bibinfo {volume}
  {15}},\ \bibinfo {pages} {1297} (\bibinfo {year} {2019})}\BibitemShut
  {NoStop}%
\bibitem [{\citenamefont {Cross}\ and\ \citenamefont
  {Hohenberg}(1993)}]{cross_pattern_1993}%
  \BibitemOpen
  \bibfield  {author} {\bibinfo {author} {\bibfnamefont {M.~C.}\ \bibnamefont
  {Cross}}\ and\ \bibinfo {author} {\bibfnamefont {P.~C.}\ \bibnamefont
  {Hohenberg}},\ }\bibfield  {title} {\bibinfo {title} {Pattern formation
  outside of equilibrium},\ }\href {https://doi.org/10.1103/RevModPhys.65.851}
  {\bibfield  {journal} {\bibinfo  {journal} {Rev. Mod. Phys.}\ }\textbf
  {\bibinfo {volume} {65}},\ \bibinfo {pages} {851} (\bibinfo {year}
  {1993})}\BibitemShut {NoStop}%
\bibitem [{\citenamefont {Goodrich}\ \emph {et~al.}(2021)\citenamefont
  {Goodrich}, \citenamefont {King}, \citenamefont {Schoenholz}, \citenamefont
  {Cubuk},\ and\ \citenamefont {Brenner}}]{goodrich_designing_2021}%
  \BibitemOpen
  \bibfield  {author} {\bibinfo {author} {\bibfnamefont {C.~P.}\ \bibnamefont
  {Goodrich}}, \bibinfo {author} {\bibfnamefont {E.~M.}\ \bibnamefont {King}},
  \bibinfo {author} {\bibfnamefont {S.~S.}\ \bibnamefont {Schoenholz}},
  \bibinfo {author} {\bibfnamefont {E.~D.}\ \bibnamefont {Cubuk}},\ and\
  \bibinfo {author} {\bibfnamefont {M.~P.}\ \bibnamefont {Brenner}},\
  }\bibfield  {title} {\bibinfo {title} {Designing self-assembling kinetics
  with differentiable statistical physics models},\ }\href
  {https://doi.org/10.1073/pnas.2024083118} {\bibfield  {journal} {\bibinfo
  {journal} {Proc. Natl. Acad. Sci. U.S.A.}\ }\textbf {\bibinfo {volume}
  {118}},\ \bibinfo {pages} {e2024083118} (\bibinfo {year} {2021})}\BibitemShut
  {NoStop}%
\bibitem [{\citenamefont {Stachowiak}\ \emph {et~al.}(2013)\citenamefont
  {Stachowiak}, \citenamefont {Brodsky},\ and\ \citenamefont
  {Miller}}]{stachowiak_costbenefit_2013}%
  \BibitemOpen
  \bibfield  {author} {\bibinfo {author} {\bibfnamefont {J.~C.}\ \bibnamefont
  {Stachowiak}}, \bibinfo {author} {\bibfnamefont {F.~M.}\ \bibnamefont
  {Brodsky}},\ and\ \bibinfo {author} {\bibfnamefont {E.~A.}\ \bibnamefont
  {Miller}},\ }\bibfield  {title} {\bibinfo {title} {A cost–benefit analysis
  of the physical mechanisms of membrane curvature},\ }\href
  {https://doi.org/10.1038/ncb2832} {\bibfield  {journal} {\bibinfo  {journal}
  {Nat Cell Biol}\ }\textbf {\bibinfo {volume} {15}},\ \bibinfo {pages} {1019}
  (\bibinfo {year} {2013})}\BibitemShut {NoStop}%
\bibitem [{\citenamefont {Falkenburger}\ \emph {et~al.}(2010)\citenamefont
  {Falkenburger}, \citenamefont {Jensen}, \citenamefont {Dickson},
  \citenamefont {Suh},\ and\ \citenamefont
  {Hille}}]{falkenburger_symposium_2010}%
  \BibitemOpen
  \bibfield  {author} {\bibinfo {author} {\bibfnamefont {B.~H.}\ \bibnamefont
  {Falkenburger}}, \bibinfo {author} {\bibfnamefont {J.~B.}\ \bibnamefont
  {Jensen}}, \bibinfo {author} {\bibfnamefont {E.~J.}\ \bibnamefont {Dickson}},
  \bibinfo {author} {\bibfnamefont {B.-C.}\ \bibnamefont {Suh}},\ and\ \bibinfo
  {author} {\bibfnamefont {B.}~\bibnamefont {Hille}},\ }\bibfield  {title}
  {\bibinfo {title} {{SYMPOSIUM} {REVIEW}: {Phosphoinositides}: lipid
  regulators of membrane proteins: {Phosphoinositides} instruct membrane
  proteins},\ }\href {https://doi.org/10.1113/jphysiol.2010.192153} {\bibfield
  {journal} {\bibinfo  {journal} {The Journal of Physiology}\ }\textbf
  {\bibinfo {volume} {588}},\ \bibinfo {pages} {3179} (\bibinfo {year}
  {2010})}\BibitemShut {NoStop}%
\bibitem [{\citenamefont {Hsieh}\ \emph {et~al.}(2021)\citenamefont {Hsieh},
  \citenamefont {Lopez}, \citenamefont {Black}, \citenamefont {Osinski},
  \citenamefont {Pawłowski}, \citenamefont {Tomchick}, \citenamefont {Liou},\
  and\ \citenamefont {Tagliabracci}}]{hsieh_dynamic_2021}%
  \BibitemOpen
  \bibfield  {author} {\bibinfo {author} {\bibfnamefont {T.-S.}\ \bibnamefont
  {Hsieh}}, \bibinfo {author} {\bibfnamefont {V.~A.}\ \bibnamefont {Lopez}},
  \bibinfo {author} {\bibfnamefont {M.~H.}\ \bibnamefont {Black}}, \bibinfo
  {author} {\bibfnamefont {A.}~\bibnamefont {Osinski}}, \bibinfo {author}
  {\bibfnamefont {K.}~\bibnamefont {Pawłowski}}, \bibinfo {author}
  {\bibfnamefont {D.~R.}\ \bibnamefont {Tomchick}}, \bibinfo {author}
  {\bibfnamefont {J.}~\bibnamefont {Liou}},\ and\ \bibinfo {author}
  {\bibfnamefont {V.~S.}\ \bibnamefont {Tagliabracci}},\ }\bibfield  {title}
  {\bibinfo {title} {Dynamic remodeling of host membranes by self-organizing
  bacterial effectors},\ }\href {https://doi.org/10.1126/science.aay8118}
  {\bibfield  {journal} {\bibinfo  {journal} {Science}\ }\textbf {\bibinfo
  {volume} {372}},\ \bibinfo {pages} {935} (\bibinfo {year}
  {2021})}\BibitemShut {NoStop}%
\bibitem [{\citenamefont {Tan}\ \emph {et~al.}(2020)\citenamefont {Tan},
  \citenamefont {Liu}, \citenamefont {Miller}, \citenamefont {Tekant},
  \citenamefont {Dunkel},\ and\ \citenamefont {Fakhri}}]{tan_topological_2020}%
  \BibitemOpen
  \bibfield  {author} {\bibinfo {author} {\bibfnamefont {T.~H.}\ \bibnamefont
  {Tan}}, \bibinfo {author} {\bibfnamefont {J.}~\bibnamefont {Liu}}, \bibinfo
  {author} {\bibfnamefont {P.~W.}\ \bibnamefont {Miller}}, \bibinfo {author}
  {\bibfnamefont {M.}~\bibnamefont {Tekant}}, \bibinfo {author} {\bibfnamefont
  {J.}~\bibnamefont {Dunkel}},\ and\ \bibinfo {author} {\bibfnamefont
  {N.}~\bibnamefont {Fakhri}},\ }\bibfield  {title} {\bibinfo {title}
  {Topological turbulence in the membrane of a living cell},\ }\href
  {https://doi.org/10.1038/s41567-020-0841-9} {\bibfield  {journal} {\bibinfo
  {journal} {Nat. Phys.}\ }\textbf {\bibinfo {volume} {16}},\ \bibinfo {pages}
  {657} (\bibinfo {year} {2020})}\BibitemShut {NoStop}%
\bibitem [{\citenamefont {Wigbers}\ \emph {et~al.}(2021)\citenamefont
  {Wigbers}, \citenamefont {Tan}, \citenamefont {Brauns}, \citenamefont {Liu},
  \citenamefont {Swartz}, \citenamefont {Frey},\ and\ \citenamefont
  {Fakhri}}]{wigbers_hierarchy_2021}%
  \BibitemOpen
  \bibfield  {author} {\bibinfo {author} {\bibfnamefont {M.~C.}\ \bibnamefont
  {Wigbers}}, \bibinfo {author} {\bibfnamefont {T.~H.}\ \bibnamefont {Tan}},
  \bibinfo {author} {\bibfnamefont {F.}~\bibnamefont {Brauns}}, \bibinfo
  {author} {\bibfnamefont {J.}~\bibnamefont {Liu}}, \bibinfo {author}
  {\bibfnamefont {S.~Z.}\ \bibnamefont {Swartz}}, \bibinfo {author}
  {\bibfnamefont {E.}~\bibnamefont {Frey}},\ and\ \bibinfo {author}
  {\bibfnamefont {N.}~\bibnamefont {Fakhri}},\ }\bibfield  {title} {\bibinfo
  {title} {A hierarchy of protein patterns robustly decodes cell shape
  information},\ }\href {https://doi.org/10.1038/s41567-021-01164-9} {\bibfield
   {journal} {\bibinfo  {journal} {Nat. Phys.}\ }\textbf {\bibinfo {volume}
  {17}},\ \bibinfo {pages} {578} (\bibinfo {year} {2021})}\BibitemShut
  {NoStop}%
\end{thebibliography}%


\begin{thebibliography}{5}%
\makeatletter
\providecommand \@ifxundefined [1]{%
 \@ifx{#1\undefined}
}%
\providecommand \@ifnum [1]{%
 \ifnum #1\expandafter \@firstoftwo
 \else \expandafter \@secondoftwo
 \fi
}%
\providecommand \@ifx [1]{%
 \ifx #1\expandafter \@firstoftwo
 \else \expandafter \@secondoftwo
 \fi
}%
\providecommand \natexlab [1]{#1}%
\providecommand \enquote  [1]{``#1''}%
\providecommand \bibnamefont  [1]{#1}%
\providecommand \bibfnamefont [1]{#1}%
\providecommand \citenamefont [1]{#1}%
\providecommand \href@noop [0]{\@secondoftwo}%
\providecommand \href [0]{\begingroup \@sanitize@url \@href}%
\providecommand \@href[1]{\@@startlink{#1}\@@href}%
\providecommand \@@href[1]{\endgroup#1\@@endlink}%
\providecommand \@sanitize@url [0]{\catcode `\\12\catcode `\$12\catcode
  `\&12\catcode `\#12\catcode `\^12\catcode `\_12\catcode `\%12\relax}%
\providecommand \@@startlink[1]{}%
\providecommand \@@endlink[0]{}%
\providecommand \url  [0]{\begingroup\@sanitize@url \@url }%
\providecommand \@url [1]{\endgroup\@href {#1}{\urlprefix }}%
\providecommand \urlprefix  [0]{URL }%
\providecommand \Eprint [0]{\href }%
\providecommand \doibase [0]{https://doi.org/}%
\providecommand \selectlanguage [0]{\@gobble}%
\providecommand \bibinfo  [0]{\@secondoftwo}%
\providecommand \bibfield  [0]{\@secondoftwo}%
\providecommand \translation [1]{[#1]}%
\providecommand \BibitemOpen [0]{}%
\providecommand \bibitemStop [0]{}%
\providecommand \bibitemNoStop [0]{.\EOS\space}%
\providecommand \EOS [0]{\spacefactor3000\relax}%
\providecommand \BibitemShut  [1]{\csname bibitem#1\endcsname}%
\let\auto@bib@innerbib\@empty
\bibitem [{\citenamefont {Mao}\ \emph {et~al.}(2019)\citenamefont {Mao},
  \citenamefont {Kuldinow}, \citenamefont {Haataja},\ and\ \citenamefont
  {Košmrlj}}]{mao_phase_2019}%
  \BibitemOpen
  \bibfield  {author} {\bibinfo {author} {\bibfnamefont {S.}~\bibnamefont
  {Mao}}, \bibinfo {author} {\bibfnamefont {D.}~\bibnamefont {Kuldinow}},
  \bibinfo {author} {\bibfnamefont {M.~P.}\ \bibnamefont {Haataja}},\ and\
  \bibinfo {author} {\bibfnamefont {A.}~\bibnamefont {Košmrlj}},\ }\bibfield
  {title} {\bibinfo {title} {Phase behavior and morphology of multicomponent
  liquid mixtures},\ }\href {https://doi.org/10.1039/C8SM02045K} {\bibfield
  {journal} {\bibinfo  {journal} {Soft Matter}\ }\textbf {\bibinfo {volume}
  {15}},\ \bibinfo {pages} {1297} (\bibinfo {year} {2019})}\BibitemShut
  {NoStop}%
\bibitem [{\citenamefont {Thiele}\ \emph {et~al.}(2013)\citenamefont {Thiele},
  \citenamefont {Archer}, \citenamefont {Robbins}, \citenamefont {Gomez},\ and\
  \citenamefont {Knobloch}}]{thiele_localized_2013}%
  \BibitemOpen
  \bibfield  {author} {\bibinfo {author} {\bibfnamefont {U.}~\bibnamefont
  {Thiele}}, \bibinfo {author} {\bibfnamefont {A.~J.}\ \bibnamefont {Archer}},
  \bibinfo {author} {\bibfnamefont {M.~J.}\ \bibnamefont {Robbins}}, \bibinfo
  {author} {\bibfnamefont {H.}~\bibnamefont {Gomez}},\ and\ \bibinfo {author}
  {\bibfnamefont {E.}~\bibnamefont {Knobloch}},\ }\bibfield  {title} {\bibinfo
  {title} {Localized states in the conserved {Swift}-{Hohenberg} equation with
  cubic nonlinearity},\ }\href@noop {} {\bibfield  {journal} {\bibinfo
  {journal} {PHYSICAL REVIEW E}\ ,\ \bibinfo {pages} {19}} (\bibinfo {year}
  {2013})}\BibitemShut {NoStop}%
\bibitem [{\citenamefont {Matthews}\ and\ \citenamefont
  {Cox}(2000)}]{matthews_pattern_2000}%
  \BibitemOpen
  \bibfield  {author} {\bibinfo {author} {\bibfnamefont {P.~C.}\ \bibnamefont
  {Matthews}}\ and\ \bibinfo {author} {\bibfnamefont {S.~M.}\ \bibnamefont
  {Cox}},\ }\bibfield  {title} {\bibinfo {title} {Pattern formation with a
  conservation law},\ }\href {https://doi.org/10.1088/0951-7715/13/4/317}
  {\bibfield  {journal} {\bibinfo  {journal} {Nonlinearity}\ }\textbf {\bibinfo
  {volume} {13}},\ \bibinfo {pages} {1293} (\bibinfo {year}
  {2000})}\BibitemShut {NoStop}%
\bibitem [{\citenamefont {Cox}(2004)}]{cox_envelope_2004}%
  \BibitemOpen
  \bibfield  {author} {\bibinfo {author} {\bibfnamefont {S.~M.}\ \bibnamefont
  {Cox}},\ }\bibfield  {title} {\bibinfo {title} {The envelope of a
  one-dimensional pattern in the presence of a conserved quantity},\ }\href
  {https://doi.org/10.1016/j.physleta.2004.10.038} {\bibfield  {journal}
  {\bibinfo  {journal} {Physics Letters A}\ }\textbf {\bibinfo {volume}
  {333}},\ \bibinfo {pages} {91} (\bibinfo {year} {2004})}\BibitemShut
  {NoStop}%
\bibitem [{\citenamefont {Cooley}\ \emph {et~al.}(1969)\citenamefont {Cooley},
  \citenamefont {Lewis},\ and\ \citenamefont {Welch}}]{cooley_fast_1969}%
  \BibitemOpen
  \bibfield  {author} {\bibinfo {author} {\bibfnamefont {J.~W.}\ \bibnamefont
  {Cooley}}, \bibinfo {author} {\bibfnamefont {P.~A.~W.}\ \bibnamefont
  {Lewis}},\ and\ \bibinfo {author} {\bibfnamefont {P.~D.}\ \bibnamefont
  {Welch}},\ }\bibfield  {title} {\bibinfo {title} {The {Fast} {Fourier}
  {Transform} and {Its} {Applications}},\ }\href@noop {} {\bibfield  {journal}
  {\bibinfo  {journal} {IEEE TRANSACTIONS ON EDUCATION}\ } (\bibinfo {year}
  {1969})}\BibitemShut {NoStop}%
\end{thebibliography}%
\end{document}


\title{
    Supplemental Information: Pattern formation of phase-separate lipid domains in bilayer membranes
}

\author{Qiwei Yu}
\affiliation{Department of Mechanical and Aerospace Engineering, Princeton University, Princeton, NJ 08544}
\affiliation{Lewis-Sigler Institute for Integrative Genomics, Princeton University, Princeton, NJ 08544}

\author{Andrej Ko\v{s}mrlj}
\affiliation{Department of Mechanical and Aerospace Engineering, Princeton University, Princeton, NJ 08544}
\affiliation{Princeton Materials Institute, Princeton University, Princeton, NJ 08544}

\maketitle
\tableofcontents
    
\section{Analytical solution of the model}
\subsection{Dynamical equations for deformation and composition}
We start with the full free energy
\begin{align}
    \Fcal[h,\phi_1,\phi_2]
     & = \Fcal_{e}[h,\phi_1] + \Fcal_c[\phi_1,\phi_2] = \int\qty[\frac{1}{2}\sigma \qty(\nabla h)^2 + \frac{1}{2}\kappa\qty(\nabla^2 h- c_1\phi_1)^2]\dbsr  \nonumber \\
     & + \int \qty[\sum_{i=1}^3\phi_i\ln\phi_i+\sum_{i\neq j}\chi_{i,j}\qty(\phi_i\phi_j-\lambda^2\nabla\phi_i\cdot\nabla\phi_j)+\xi\phi_1\phi_2\phi_3]\dbsr,
\end{align}
$h$ follows Model A dynamics (not conserved) and $\phi_{1,2}$ follows Model B dynamics (conserved):
\begin{align}
    \pdv{h}{t}     = -M_h\frac{\delta F}{\delta h}, \quad
    \pdv{\phi_i}{t}  = M_i \nabla^2\frac{\delta F}{\delta \phi_i}.
\end{align}
$\phi_3=1-\phi_1-\phi_2$ is dependent on $\phi_{1,2}$ by incompressibility.
As mentioned in the main text, space and time are measured in $\lambda$ and $nk_BT M/\lambda^2$ respectively, and energy is measured in $nk_BT$. Hence, we set $M=1$ and $\lambda=1$ in the following derivation.
The equations of motion are
\begin{align}
    \frac{\partial h}{\partial t}      & = \sigma \nabla^2 h - \kappa \nabla^4 h + \kappa c_1 \nabla^2\phi_1,  \label{Eq:dhdt}\\
    \frac{\partial \phi_1}{\partial t} & = \kappa c_1^2 \nabla^2\phi_1 - \kappa c_1 \nabla^4h + \nabla^2\frac{\delta\Fcal_c}{\delta \phi_{1}}, \label{Eq:dphi1dt}\\
    \frac{\partial \phi_2}{\partial t} & =  \nabla^2\frac{\delta\Fcal_c}{\delta \phi_{2}}. \label{Eq:dphi2dt}
\end{align}
where $\frac{\delta\Fcal_c}{\delta \phi_{1,2}}$ are the chemical potentials due to the (generalized) Flory-Huggins free energy:
\begin{align}
    \frac{\delta\Fcal_c}{\delta \phi_{1}} & = \ln\left(\frac{\phi_1}{1-\phi_1-\phi_2}\right) + (\chi_{12}-\chi_{23})(1+\nabla^2)\phi_2 + \chi_{13}(1+\nabla^2)(1-2\phi_1-\phi_2)+\xi\phi_2(1-2\phi_1-\phi_2), \\
    \frac{\delta\Fcal_c}{\delta \phi_{2}} & = \ln\left(\frac{\phi_2}{1-\phi_1-\phi_2}\right) + (\chi_{12}-\chi_{13})(1+\nabla^2)\phi_1 + \chi_{23}(1+\nabla^2)(1-\phi_1-2\phi_2)+\xi\phi_1(1-\phi_1-2\phi_2).
\end{align}

\subsection{Pattern size and morphology}
Fourier transforming Eq.~\eqref{Eq:dhdt} yields the steady-state solution for $h$:
\begin{align}
    h(\bsk) =- \frac{\kappa c_1}{\kappa k^2+\sigma} \phi_1(\bsk).
\end{align}
Let $\delta \phi_i=\phi_i-\bar{\phi}_i$. The free energy for membrane deformation reduces to
\begin{align}
    \Fcal_e[h,\phi_1] 
    & =  \int\qty[\frac{1}{2}\sigma \qty(\nabla h)^2 + \frac{1}{2}\kappa\qty(\nabla^2 h- c_1\bar{\phi_1}-c_1\delta\phi_1)^2]\dbsr  \\
    & = \frac{A}{2}\kappa c_1^2\bar{\phi_1}^2 + \int 
    \dk\qty[
        \frac{1}{2}(\sigma+\kappa k^2)k^2 h(\bsk)h(-\bsk) + \frac{1}{2}\kappa c_1^2 \delta\phi_1(\bsk)\delta\phi_1(-\bsk) 
        + \kappa c_1 k^2 \bar{\phi_1}\delta\phi_1(-\bsk)h(\bsk)
    ] \\
    & =  \frac{A}{2}\kappa c_1^2\bar{\phi_1}^2 + \frac{1}{2} \kappa c_1^2 \int\frac{\abs{\delta \phi_1(\bsk)}^2}{1+ \frac{\kappa k^2}{\sigma}}\dk, \label{Eq:fe}
\end{align} 
where $A$ is the total area of the membrane. 

Assuming phase separation along the tie lines of $\Fcal_c$ (which can be determined by the convex hull construction~\cite{mao_phase_2019}), the ratio of the concentration change of the two phospholipid species: $s=-\delta \phi_2/\delta \phi_1$. 

\begin{align}
    \Fcal_c[\phi_1,\phi_2]
    =                              & 
    A f_c(\bar \phi_1, \bar\phi_2) +
    \int \qty[-\frac{a}{2} \delta\phi_1^2 + \frac{c}{3} \delta \phi_1^3 + \frac{b}{4} \delta \phi_1^4 + \frac{1}{2}\mu \qty(\nabla\delta \phi_1)^2] \dbsr,\label{Eq:fc}
\end{align} 
The expansion parameters are:
\begin{align}
    \mu = & 2\qty(s\chi_{12}-(s-1)\chi_{13}+ s(s-1)\chi_{23}),                   \label{EqS:mu}                                      \\
    a =   & -\qty(\bar{\phi}_1^{-1}+s^2\bar{\phi}_2^{-1}+(s-1)^2\bar{\phi}_3^{-1})+\mu + 2\xi\qty[s(s-1)\bar{\phi}_1 - (s-1)\bar{\phi}_2 + s\bar{\phi}_3], \\
    c =   & \frac{1}{2}\qty(-\bar{\phi}_1^{-2} + s^3 \bar{\phi}_2^{-2}-(s-1)^3 \bar{\phi}_3^{-2} - 6s(s-1)\xi),                          \\
    b =   & \frac{1}{3}\qty(\bar{\phi}_1^{-3} + s^4 \bar{\phi}_2^{-3} + (s-1)^4 \bar{\phi}_3^{-3}).
\end{align}
Here, $\mu$ is related to the line tension between the two phases. In the absence of membrane deformation ($h=0$), phase separation requires $a>0$. Note that $b>0$ ensures that $\delta \phi_1$ remains bounded. 

Combining Eq.~\eqref{Eq:fe} and Eq.~\eqref{Eq:fc} leads to the total free energy given by Eq.~(7) of the main text (up to a constant which can be removed by shifting the zero of the free energy):
\begin{align}
    \mathcal{F}_\mathrm{eff}[\delta \phi_1]=\frac{1}{2} \int \qty(\frac{\sigma\kappa c_1^2}{\sigma+ \kappa k^2} + \mu k^2) \abs{\delta\phi_1(\bsk)}^2\dd[2]{\bsk}
    +\int \qty[-\frac{a}{2}\delta\phi_1(\bsr)^2 + \frac{c}{3} \delta\phi_1(\bsr)^3 + \frac{b}{4} \delta\phi_1(\bsr)^4]\dbsr. \label{Eq:feff}
\end{align}
Minimizing the free energy with respect to $\abs{\bsk}$ leads to $k^2=k_c^2=\sqrt{\frac{\sigma c_1^2}{\mu}}-\frac{\sigma}{\kappa}$ (Eq.~(8) of the main text). Expanding the free energy around $k_c$: 
\begin{align}
    \frac{\sigma\kappa c_1^2}{\sigma+ \kappa k^2} + \mu k^2
    = 2\sqrt{\mu\sigma c_1^2} - \frac{\sigma\mu}{\kappa} + \sqrt{\frac{\mu^3}{\sigma c_1^2}}\qty(k^2-k_c^2)^2 + O\qty[(k^2-k_c^2)^3].
\end{align}
Thus, the dynamical equation for $\phi_1$ is 
\begin{align}\label{Eq:phi1}
    \pdv{\delta\phi_1}{t} = \nabla^2 \frac{\delta \Fcal}{\delta \phi_1} 
    = \nabla^2\qty[\qty(-\aeff+\sqrt{\frac{\mu^3}{\sigma c_1^2}}\qty(k_c^2+\nabla^2)^2)\delta\phi_1 + c  \delta\phi_1^2 + b \delta\phi_1^3 ],
\end{align}
where 
\begin{align}\label{Eq:aeff}
    \aeff = a - 2\sqrt{\mu\sigma c_1^2} + \frac{\sigma\mu}{\kappa}.
\end{align}
Eq.~\eqref{Eq:phi1} has the same form as the conserved (or derivative) Swift-Hohenberg equation~\cite{thiele_localized_2013,matthews_pattern_2000,cox_envelope_2004}. 
Applying amplitude expansion $\delta\phi_1 = \phi_1-\bar{\phi_1} = \sum_n A_{n} e^{i\bsk_n\cdot\bsr} + \mathrm{c.c.}$ with $\abs{k_n}=k_c$ and $n=3$ (i.e., the wavevectors $\bs{k}_{1,2,3}$ are of the same magnitude and oriented at 120 degrees with respect to each other) leads to the amplitude equations:
\begin{align}
    \dv{A_1}{\tau} & = \qty(\aeff -3b \abs{A_1}^2-6b\abs{A_2}^2-6b\abs{A_3}^2)A_1 - 2cA_2^*A_3^*,  \\
    \dv{A_2}{\tau} & = \qty(\aeff -3b \abs{A_2}^2-6b\abs{A_1}^2-6b\abs{A_3}^2)A_2 - 2cA_1^*A_3^*, \\
    \dv{A_3}{\tau} & = \qty(\aeff -3b \abs{A_3}^2-6b\abs{A_1}^2-6b\abs{A_2}^2)A_3 - 2cA_1^*A_2^*.
\end{align}
where $\tau = k_c^2t$. Let $A_i = R_i \exp(i\theta_i)$. The dynamics of the amplitude $R$ and phase $\theta$ are given by 
\begin{align}
    \frac{\dot{R_1}}{R_1} = \dv{\ln R_1}{\tau} &= \mathrm{Re}\qty{\dv{\ln A_1}{\tau}} = \aeff -3b R_1^2 - 6b R_2^2 - 6b R_3^2 - 2c \frac{R_2 R_3}{R_1} \cos(\theta_1+ \theta_2+\theta_3), \\
    \dot{\theta_1} =\dv{\theta_1}{\tau} &= \mathrm{Im}\qty{\dv{\ln A_1}{\tau}} = 2c \frac{R_2 R_3}{R_1} \sin(\theta_1+ \theta_2+\theta_3).
\end{align}
The sum of phases $\Theta = \theta_1+ \theta_2+\theta_3$ evolves following
\begin{align}
    \dv{\Theta}{\tau} = 2cQ\sin\Theta,\quad
    \text{where\ } Q = \sum_{\mathrm{cyc.}} \frac{R_{i+1}R_{i+2}}{R_i}>0,
\end{align}
where $\mathrm{cyc.}$ represents cyclic summation over indices $i=1,2,3$. 
Thus, $\Theta$ has two fixed points $\Theta=0$ and $\Theta=\pi$. However, only the one corresponding to $\cos\Theta = -\mathrm{sgn}(c)$ is stable. Substituting it to the amplitude equation for $R_1$ leads to
\begin{align}
    \dv{R_1}{\tau} &= R_1\qty(\aeff -3b R_1^2 - 6b R_2^2 - 6b R_3^2 + 2\abs{c} \frac{R_2 R_3}{R_1}).
\end{align}
The fixed points of this equation represents different patterns. To analyze the stability of the fixed points, we also compute the Jacobian:
\begin{align}
    J = \pdv{\dot{R}}{R} = \begin{pmatrix}
    \aeff - 9b R_1^2 - 6b R_2^2 - 6b R_3^2 & -12b R_1 R_2 + 2\abs{c} R_3 & -12b R_1 R_3 + 2\abs{c} R_2\\
    -12b R_1 R_2 + 2\abs{c} R_3 & \aeff - 9b R_2^2 - 6b R_1^2 - 6b R_3^2 & -12b R_2 R_3 + 2\abs{c} R_1\\
    -12b R_1 R_3 + 2\abs{c} R_2 & -12b R_2 R_3 + 2\abs{c} R_1 & \aeff - 9b R_3^2 - 6b R_1^2 - 6b R_2^2
    \end{pmatrix}.
\end{align}

This equation has three fixed points:
\begin{itemize}
    \item Uniform state: $R_1=R_2=R_3=0$. The fixed point is stable when $\aeff<0$.
    \item Stripe state: $R_1= \sqrt{\frac{\aeff}{3b}}$ and $R_2=R_3=0$. The Jacobian is
    \begin{align}
        J = \begin{pmatrix}
        -2\aeff & 0 & 0\\
        0 & -\aeff & 2\abs{c}\sqrt{\frac{\aeff}{3b}} \\
        0 & 2\abs{c}\sqrt{\frac{\aeff}{3b}} & -\aeff
        \end{pmatrix}.
    \end{align}
    The fixed point is stable when the eigenvalues have negative real parts, which requires
    \begin{align}
        \aeff^2 - 4c^2 \frac{\aeff}{3b} >0 \Rightarrow \frac{\aeff b}{c^2} > \frac{4}{3}.
    \end{align}
    \item Dots (hexagonal) state: $R_1=R_2=R_3=R$, which is given by
    \begin{align}
        \aeff +2\abs{c} R -15b R^2 =0 \Rightarrow R_{\pm} = \frac{\abs{c} \pm \sqrt{c^2 + 15\aeff b}}{15b}= \frac{\abs{c}}{15b}\qty(1\pm\sqrt{1+\frac{15\aeff b}{c^2}}).
    \end{align}
    The solution exists when $\aeff>-\frac{c^2}{15b}$. 
    By computing the eigenvalues of the Jacobian, it can be shown that $R_{-}$ is always an unstable fixed point, while $R_+$ is stable when $ \aeff<\frac{16c^2}{3b}$. 
\end{itemize}
Combining the above results, we find that the stability of the fixed points can be captured by a single control parameter $g = \frac{a_\mathrm{eff}b}{c^2} \equiv \frac{b}{c^2}\qty(a - 2  \sqrt{c_1^2\sigma\mu} + \frac{\mu\sigma}{\kappa})$. The uniform state is stable when $g<0$, the stripe state is stable when $g>\frac{4}{3}$, and the dots state is stable when $-\frac{1}{15}<g<\frac{16}{3}$.

\subsection{Total free energy density}
Here, we present the free energy density of the patterns, which is used in Fig.~3 of the main text.
Following Eq.~\eqref{Eq:feff}, the free energy density of the system is given by
\begin{align}
    f =  f_0-\frac{\aeff}{2}\expval{\delta \phi_1^2} + \frac{c}{3}  \expval{\delta\phi_1^3} + \frac{b}{4}  \expval{\delta\phi_1^4},
\end{align}
where $f_0$ is the free energy density of the uniform state, and $\expval{\cdot}$ represents spatial averaging over the entire system.

For the stripe state, we have:
\begin{align}
    \expval{\phi^2} = 2R^2 = \frac{2\aeff}{3b},\quad
    \expval{\phi^3} = 0,\quad
    \expval{\phi^4} = 6R^4 = \frac{2\aeff^2}{3b^2}.\\
    f_\mathrm{stripe} = f_0-\frac{\aeff}{2}\frac{2\aeff}{3b} + \frac{b}{4}\frac{2\aeff^2}{3b^2}= f_0 -\frac{\aeff^2}{6b}.
\end{align}

For the dot state, we have:
\begin{align}
    \expval{\phi^2}
    = 6R^2,\quad
    \expval{\phi^3}
    = -12R^3 \mathrm{sgn}(c),\quad
    \expval{\phi^4}
    = 90R^4.\\
    f_\mathrm{dot} =  f_0-R^2\qty(
       \abs{c} R + \frac{3}{2}\aeff
    ),\quad \text{where}\quad R = \frac{\abs{c}}{15b}\qty(1+\sqrt{1+\frac{15\aeff b}{c^2}}).
\end{align}

\subsection{Validity of the tie line approximation}
\begin{figure}[bt]
    \includegraphics[width=\linewidth]{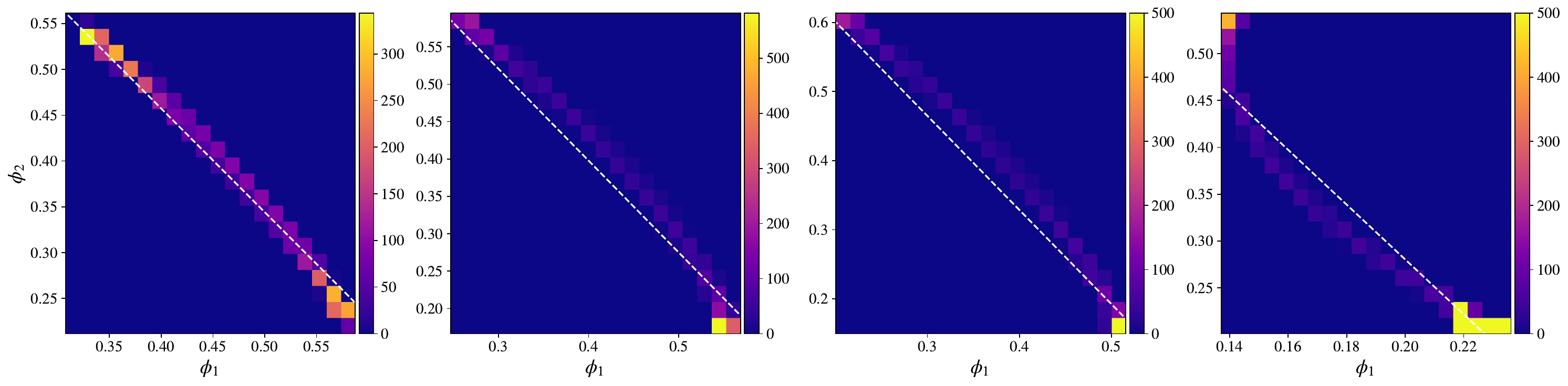}
    \caption{
        The probability density function (PDF) of the local composition $P(\phi_1,\phi_2)$ for the 4 steady-state patterns shown in Fig.~1C of the main text (from left to right are the PDFs corresponding to the upper left, upper right, lower left, and lower right panels of Fig.~1C). The white dashed lines are the tie lines obtained from the convex hull construction. 
        For the last two panels, the color bar is truncated at 500 to make the distribution visible.   
    }
    \label{sfig:tie line}
\end{figure}

In order to justify expanding the free energy along the tie line, we plot the histogram of local composition $(\phi_1(r), \phi_2(r))$ and compare it with the tie line. 
Fig.~\ref{sfig:tie line} shows the histograms for the 4 steady-state patterns shown in Fig.~1C of the main text. Indeed, the histograms are peaked around the tie line, which justifies fixing the ratio $s=-\delta \phi_2/\delta \phi_1$ to that of the tie line. 

\subsection{Using analytical $\mu$ to predict characteristic domain size}
\begin{figure}[bt]
    \includegraphics[width=0.6\linewidth]{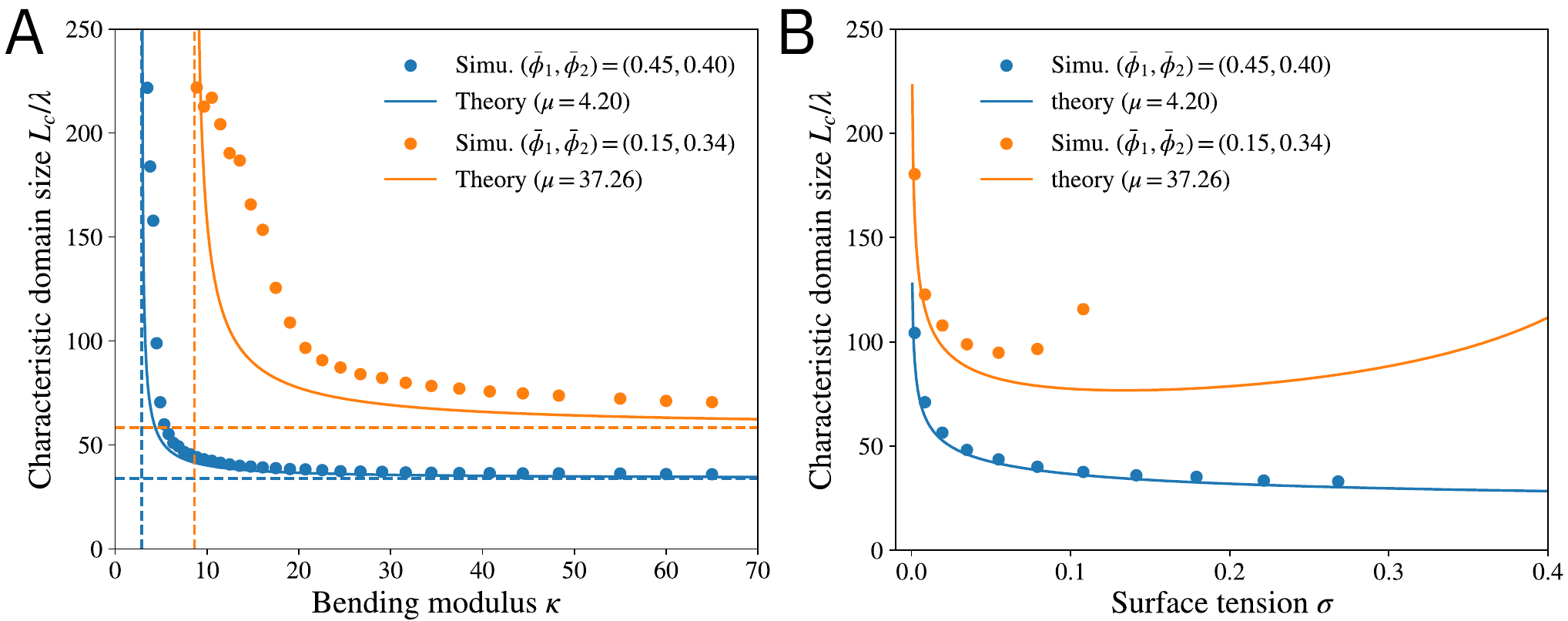}
    \caption{
        The characteristic domain size as functions of (A)~the bending modulus $\kappa$ and (B)~surface tension $\sigma$. The parameters are the same as Fig.~2C,D of the main text, except for $\mu$ which is estimated from Eq.~\eqref{EqS:mu} rather than obtained by fitting.
    }
    \label{sfig:Fig2CD}
\end{figure}

In Fig.~2C,D of the main text, the characteristic domain size was captured with a single fitting parameter $\mu$. This parameter can also be estimated from Eq.~\eqref{EqS:mu}, which leads to slightly worse agreement (see Fig.~\ref{sfig:Fig2CD}).

\section{Details of the numerical simulations}
The dynamical equations Eq.~\eqref{Eq:dhdt}--\eqref{Eq:dphi2dt} are solved in a $L\times L$ square domain with periodic boundary conditions. Space is discretized into $N\times N$ grid points, and time is discretized to steps of $\Delta t$. 
Time integration is performed in the Fourier space $y_i^n(\bs{k}) = \int y_i^n(\bs{r})\exp(-i\bs{k}\cdot\bs{r})\dd\bs{r}$ using an implicit-explicit scheme:
\begin{align}
    \frac{\bs{y}^{n+1}-\bs{y}^{n}}{\Delta t} = \bs{L}(\bs{y}^{n+1}) + \bs{N}(\bs{y}^{n}),
\end{align}
where $\bs{y}=(h, \phi_1, \phi_2)^T$ and $n$ labels the time steps. $\bs{L}(\bs{y})=\mathbf{A}\cdot \bs{y} $ is the linear and implicit part of the equation and $\bs{N}$ is the explicit part. They are converted back and forth between real space and Fourier space representations~\cite{cooley_fast_1969}.

For this study, the implicit part is diagonal: $L_i = k^2 A_i y_i$, with $A=\qty(\sigma + \kappa k^2, \kappa c_1^2 + 2\chi_{13}\lambda^2 k^2, 2\chi_{23}\lambda^2 k^2)$. All the other terms in the equation are encompassed in the explicit part $\bs{N}$. 
Thus, the update rule reads:
\begin{align}
    y_i^{n+1} = \frac{y_i^n + N_i(y_i^n)}{1+ A_i k^2 \Delta t}.
\end{align}
To improve the accuracy, we also perform $m_{pc}$ predictor-corrector iterations for each time step: 
\begin{align}
    y_i^{n+1, j+1} = \frac{y_i^n + \frac{1}{2}\qty[N_i\qty(y_i^n)+N_i\qty(y_i^{{n+1},j})]}{1+ A_i k^2 \Delta t},
\end{align}
where $j$ labels the predictor-corrector iterations with $y_i^{n+1,0}=y_i^n$ and $y_i^{n+1} = y_i^{n+1,m_{pc}}$.

The typical parameters used in this work are $L=1000\lambda$, $N=512$, $\Delta t=1$, $m_{pc}=3$.

\begin{figure}[b!]
    \includegraphics[width=0.6\linewidth]{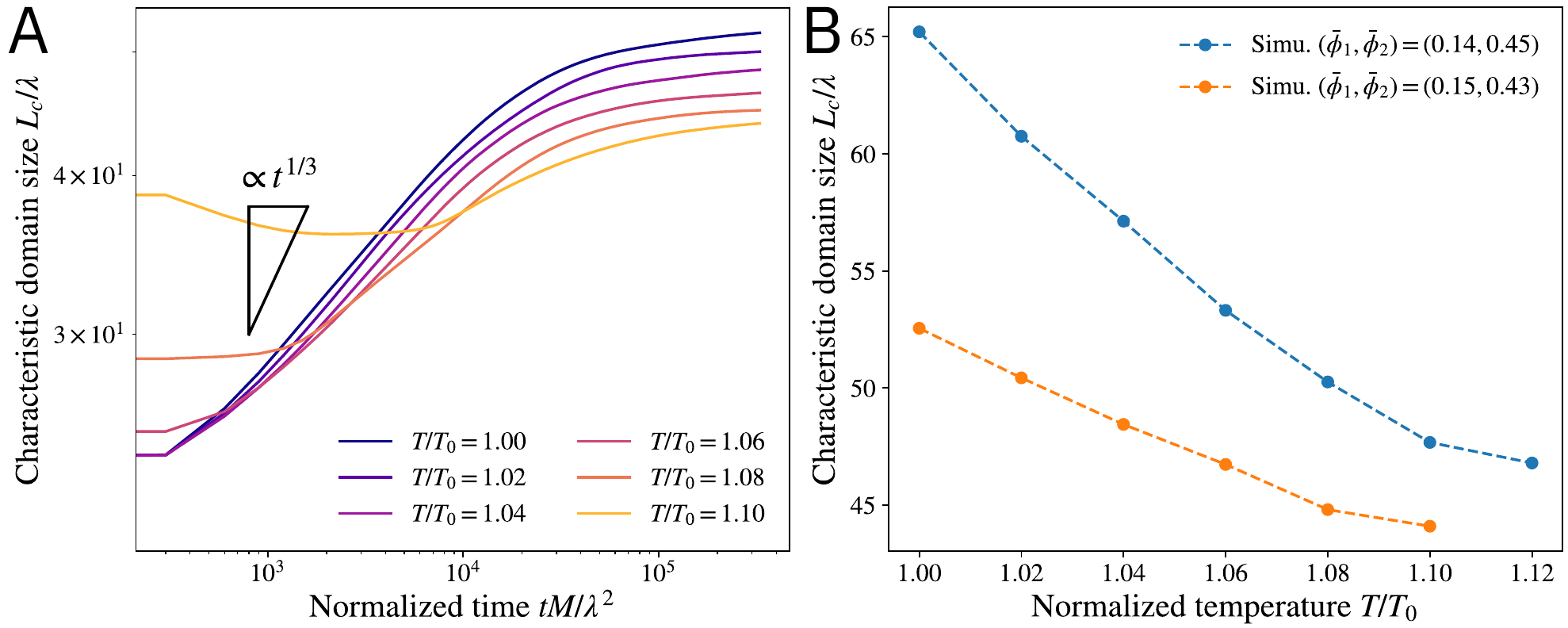}
    \caption{
    (A) The time evolution of the characteristic domain size for different temperature $T$. Mean composition: $(\bar{\phi}_1,\bar{\phi}_2)=(0.38,0.36)$. All the other parameters are the same as in Fig.~1 of the main text.
    (B) The characteristic domain size as a function of the temperature $T$.
    }
    \label{sfig:temperature Lc}
\end{figure}

\begin{figure}[bt]
    \includegraphics[width=\linewidth]{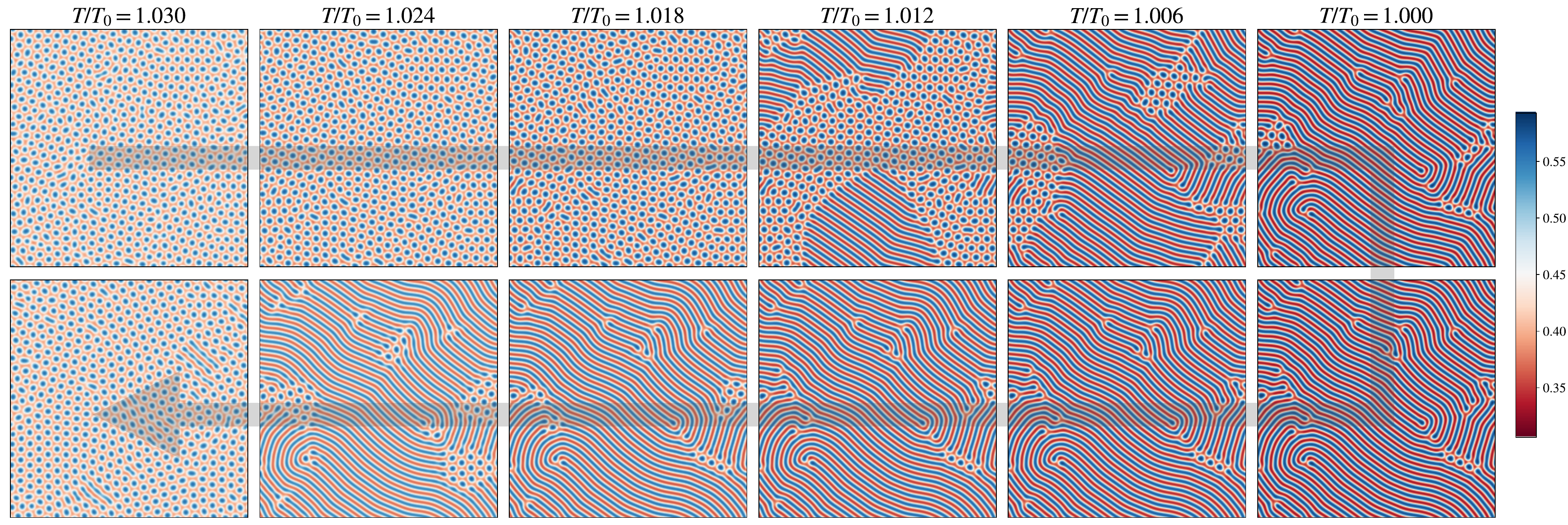}
    \caption{Pattern hysteresis with respect to temperature. The temperature is first decreased and then increased according to the black arrow. Mean composition: $(\bar{\phi}_1,\bar{\phi}_2)=(0.45,0.40)$. All the other parameters are the same as in Fig.~1 of the main text.}
    \label{sfig:temperature hysteresis}
\end{figure}

\section{Temperature dependence of the patterns}

Here, we assume that the material properties and the interaction energies do not vary strongly with temperature. Thus, the temperature dependence predominantly enters through the entropy term. 
Since energy is measured in units of $nk_BT$, the rescaled parameters $(\bar{\sigma}, \bar{\kappa},\chi, \xi)$ scale with temperature with a factor of $1/(k_BT)$. In the simulations, we tune temperature by rescaling these parameters by a factor of $\frac{T_0}{T}$, where $T_0$ is the reference temperature.
Fig.~\ref{sfig:temperature Lc} shows that the coarsening is arrested at a finite length scale (left), which decreases with temperature (right). 
Fig.~\ref{sfig:temperature hysteresis} shows the hysteresis of the pattern morphology with respect to temperature.
As the temperature is decreased, the pattern starts to morph from the dot state to the stripe state at $T/T_0\approx 1.012$, while the backward transition does not occur until $T/T_0\approx 1.024$.

\bibliography{membrane}